\documentclass[%
reprint,
superscriptaddress,
 amsmath,amssymb,
 aps,
floatfix,
]{revtex4-1}
\usepackage{subfigure}
\usepackage{graphicx}
\usepackage{dcolumn}
\usepackage{bm}
\usepackage[super]{nth}
\usepackage{mathtools}
\usepackage{lipsum}

\begin{document}

\title{Magnetic interactions in AB-stacked kagome lattices: magnetic structure, symmetry, and duality}

\author{A. Zelenskiy}
 \affiliation{Department of Physics and Atmospheric Science, Dalhousie University, Halifax, Nova Scotia, Canada B3H 3J5}

\author{T. L. Monchesky}
\affiliation{Department of Physics and Atmospheric Science, Dalhousie University, Halifax, Nova Scotia, Canada B3H 3J5}%

\author{M. L. Plumer}
 \affiliation{Department of Physics and Atmospheric Science, Dalhousie University, Halifax, Nova Scotia, Canada B3H 3J5}
 \affiliation{Department of Physics and Physical Oceanography, Memorial University of Newfoundland, St. John’s, Newfoundland, A1B 3X7, Canada}
 
\author{B. W. Southern}
 \affiliation{Department of Physics and Astronomy, University of Manitoba, Winnipeg, Manitoba, Canada R3T 2N2}

\date{\today}

\begin{abstract}
We present the results of an extensive study of the phase diagram and spin wave excitations for a general spin model on a hexagonal AB-stacked kagome system.
The boundaries of the magnetic phases are determined via a combination of numerical (Monte Carlo) and analytical (Luttinger-Tisza) methods.
We also determine the phase coexistence regions by considering the instabilities in the spin wave spectra. 
Depending on the strength of the spin-orbit coupling (SOC), some spin and lattice rotations become decoupled, leading to considerably larger symmetry groups than typical magnetic groups.
Thus, we provide a detailed symmetry description of the magnetic Hamiltonian with negligible, weak, and intermediate strength of SOC. 
The spin symmetry in these three cases has a strong effect on the splittings observed in the spin excitation spectra.
We further identify a number of self-duality transformations that map the Hamiltonian onto itself. 
These transformations describe the symmetry of the parameter space and provide an exact mapping between the properties of different magnetic orders and lead to accidental degeneracies.
Finally, we discuss the physical relevance of our findings in the context of $\mathrm{Mn}_3X$ compounds.
\end{abstract}

\pacs{Valid PACS appear here}
\maketitle


\section{\label{sec:intro}Introduction}

Magnetic materials on lattices comprised of equilateral triangles continue to attract attention from both the experimental and theoretical viewpoint due to the richness of physical properties provided by geometric frustration~\cite{sadoc_mosseri_frustration_1999,diep_frustration_2005,moessner_ramirez-frustration_2006,lacroix_frustration_2011}.
Frustrated antiferromagnets are often characterized by non-collinear spin configurations, since the topology of the lattice forbids a conventional N\'eel order.
Among the many realizations of frustrated systems, kagome magnets with antiferromagnetic nearest-neighbor (nn) interactions have become a staple example of systems with macroscopic degeneracy in the ground state, giving rise to high sensitivity of various symmetry-breaking perturbations~\cite{Chalker_Holdsworth_Shender_kagome_1992_prl,Harris_Kallin_Berlinsky_kagome_1992_prb,Huse_Rutenberg_kagome_1992_prb,Balents_Fisher_Girvin_kagome_2002_prb,Fu_kagome_2015_science,Fujihala_kagome_2020_nature}.

Recently, a family of magnetic materials with AB-stacked kagome layer structure and a general formula $\mathrm{Mn}_3X$ have been experimentally shown to host the anomalous Hall effect (AHE) as well as anomalous Nernst effect (ANE)~\cite{Chen_Niu_MacDonald_ahe_2014_prl,Nakatsuji_Kiyohara_Higo_ahe_2015_nature,Kiyohara_ahe_2016_prap,Nayak_ahe_2016_science,Ikhlas_ane_2017_nature,Hong_ane_2020_prm}.
These discoveries prompted recent theoretical and experimental studies of the magnetic properties in $\mathrm{Mn}_3X$ magnets~\cite{Liu_Balents_gs_2017_prl,Park_gs_2018_nature_pub,Li_gs_2019_nature_comm,Reichlova_gs_2019_nature_comm,Soh_gs_2020_prb,Chen_gs_2020_prb,Zelenskiy_Monchesky_Plumer_Southern_2021_prb}.

Over the last decade, magnets with strong SOC have been under an intense investigation motivated by an ongoing search for unconventional magnetic phases.
On the one hand, these include strongly correlated disordered states with large degeneracy in the ground states, such as various types of spin liquids~\cite{kitaev_spinliquid_2006,Ran_spinliquid_2007_prl,Balents_spinliquid_2010_nature,Yan_spinliquid_2011_science,Messio_spinliquid_2012_prl,Bauer_spinliquid_2014_nature_comm,Savary_Balents_2016_RPP,Takagi_spinliquid_2019_nature}.
On the other hand, however, ordered magnetic textures such as skyrmion lattices~\cite{Bogdanov_Hubert_skyrmions_1994_JMMM,Rossler_Bogdanov_Pfleiderer_skyrmions_2006_nature}, and multi-$\mathbf{Q}$ structures consisting of linear superpositions of non-colinear spin density waves~\cite{Hayami_frustrated_skyrmions_2021_prb,Leonov_Mostovoy_skyrmions_2015_nature}, have attracted increasing interest in the literature. These non-trivial magnetic orders often have nonzero scalar spin chirality, which serves as a source of the emergent electromagnetic fields, within the Berry phase formalism~\cite{Schulz_skyrmions_emergent_2012_nature,Everschor-Sitte_Sitte_the_2014_jap}, giving rise to important transport properties, such as topological Hall effect (THE)~\cite{Neubauer_the_2009_prl,Bruno_the_2004_prl,Everschor-Sitte_Sitte_the_2014_jap,Kurumaji_the_2019_science,He_the_2022_acta_mater} and spin Hall effect (SHE)~\cite{Hirsch_she_1999_prl,Chen_Byrnes_she_2019_prb}.
More recently, magnetic frustration was identified as one of the stabilizing factors for multi-$\mathbf{Q}$ spin configurations, leading to magnetic orders beyond those typically observed in chiral ferromagnets~\cite{Hayami_frustrated_skyrmions_2021_prb}.

In the case of $\mathrm{Mn}_3X$ compounds, both the THE and SHE have been observed experimentally, and studies have also established that both the AHE, and ANE, as well as the magnetic structure are strongly anisotropic, implying that the spin couplings beyond isotropic exchange are crucial for understanding the magnetic properties of these systems~\cite{Kiyohara_ahe_2016_prap,Nayak_ahe_2016_science,Zhang_ahe_anis_2017_prb}. 
Previous studies have also established that different types of anisotropic interactions compete with each other leading to additional frustration~\cite{Liu_Balents_gs_2017_prl,Soh_gs_2020_prb,Chen_gs_2020_prb,Zelenskiy_Monchesky_Plumer_Southern_2021_prb}. 
These facts motivate a systematic study of the magnetic ground state properties in an extended parameter space.

In our previous work in Ref.~\cite{Zelenskiy_Monchesky_Plumer_Southern_2021_prb} (hereafter referred to as Ref.~I), we derived a magnetic model for these magnetic compounds using general symmetry principles.
Apart from the typical exchange couplings, the symmetry-allowed terms consist of various anisotropic couplings, including Dzyaloshinskii-Moriya (DM), bond-dependent anisotropic exchange, and single-ion anisotropy (SIA).
These anisotropic interactions arise from the coupling of spins to the underlying lattice via SOC. 
In Ref.~I, we have studied the ground state properties and spin wave excitations of this model relevant to the magnetism of $\mathrm{Mn}_3X$ systems.
In particular, this work provided a detailed analysis of the interplay between various anisotropic interactions and their effect on the static and dynamic properties of $\mathrm{Mn}_3X$ compounds.
We determined that both SIA and bond-dependent anisotropy compete with the DM interactions, leading to an induced magnetic moment and an excitation spectrum with broken six-fold symmetry.
However, we also found that these properties are extremely sensitive to both the signs and relative magnitudes of the anisotropic interactions.

In the present work, we investigate semi-classical ground state properties of hexagonal AB-stacked kagome systems for an extended range of magnetic interactions using a combination of analytical Luttinger-Tisza (LT) and numerical Monte-Carlo (MC) techniques. 
\begin{figure}[t]
    \centering
    \includegraphics[width=0.45\textwidth]{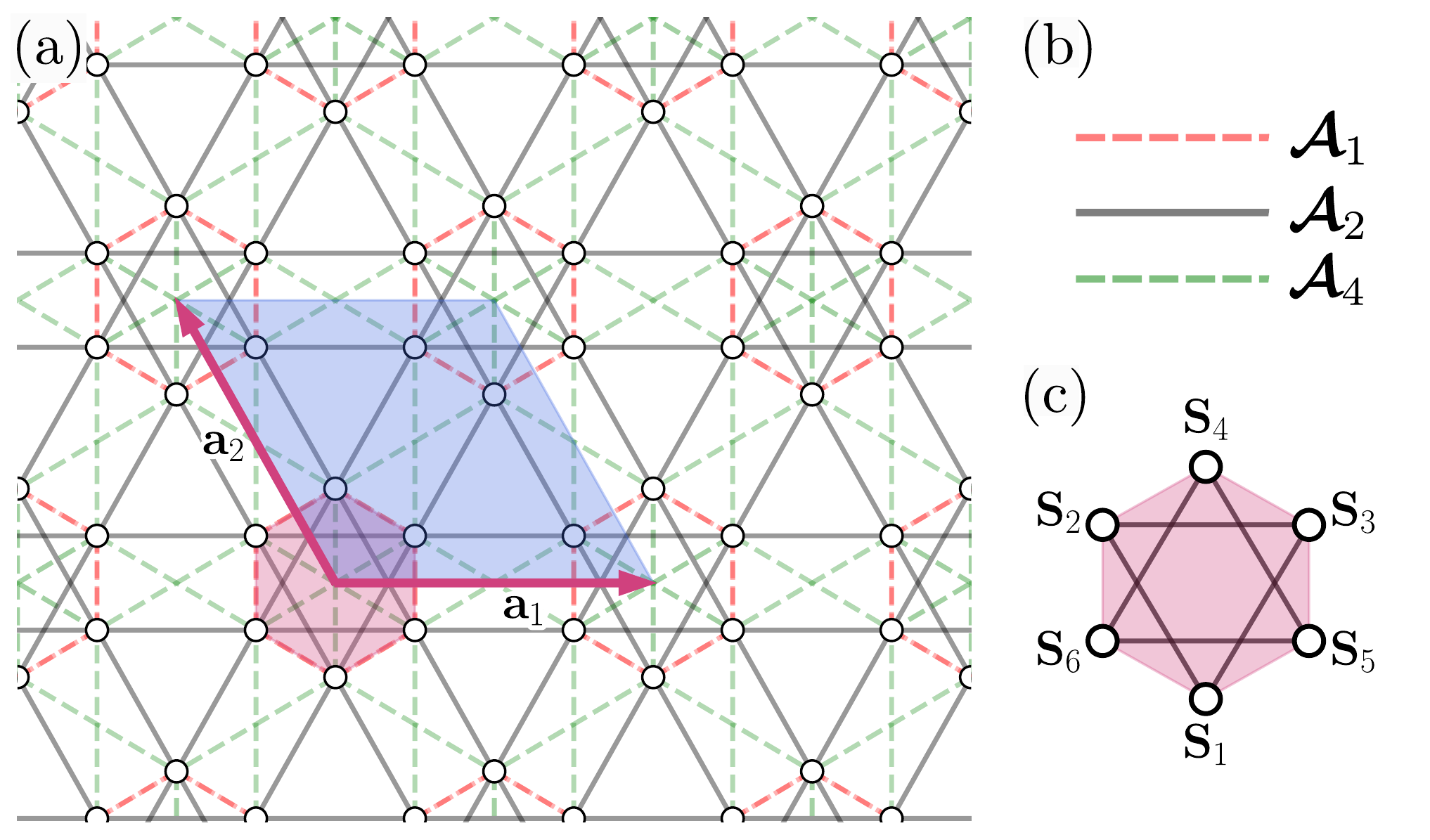}
    \caption{(a) A sketch of the AB-stacked kagome lattice. The conventional unit cell defined by the lattice vectors $\mathbf{a}_1$ and $\mathbf{a}_2$ is shaded in blue, while the hexagon formed by the four sublattices is shaded in pink. (b) The three types of nn and nnn interactions appearing in (a). Solid black lines indicate the nn in-plane interactions, while the dashed red and green lines represent the nn and nnn out-of-plane interactions respectively ($\boldsymbol{\mathcal{A}}_i = \{J_i,D_i,A_i^{(z)},A_i^{(xy)}\}$). (c) An enlarged diagram of the unit cell convention used in this work. $\mathbf{S}_i$ label the spins on the six sublattices. Spins 1, 2, and 3 reside in layer A, while spins 4, 5, and 6 are in layer B.}
    \label{fig:structure}
\end{figure}
To tackle the large parameter space of the magnetic model, we group the magnetic interactions based on the effective spin symmetry imposed by the relative strength of the SOC.
We determine three SOC symmetry regimes and present the corresponding group structure, along with the relevant irreducible representations (irreps).
Furthermore, for each of these symmetry regimes, we identify a set of self-duality transformations that reduce the number of independent points in the parameter space.
We find that in the weak SOC limit the magnetic Hamiltonian has the largest number of dualities which comprise a group with non-Abelian structure. 
The numerical and analytical calculations reveal a variety of magnetic phases, including single-$\mathbf{Q}$, multi-$\mathbf{Q}$, as well as more complicated structures with delocalized structure factors in the Brillouin zone.
We parameterize the magnetic phases and study the elementary spin wave excitations.
Finally, we analyze the effects of bond-dependent exchange and SIA on the magnetic phases stabilized by exchange and DM interactions, and discuss the implications for the case of $\mathrm{Mn}_3X$ systems.

The rest of this paper is organized as follows. 
In Sec.~\ref{sec:model_methods} we introduce the magnetic model and briefly outline the analytical and numerical methods.
In Sec.~\ref{sec:symmetry} we identify the connection between the strength of the SOC and the effective symmetry of the magnetic system.
Next, in Sec.~\ref{sec:self_duality}, the self-duality transformations are introduced and derived for each SOC symmetry case.
The magnetic ground state phase diagrams for models with exchange and DM interactions are presented in Sec.~\ref{sec:phase_diagram}.

The distinct types of magnetic order are described in Sec.~\ref{sec:phases_structure}, and the corresponding spin-wave excitation spectra are given in Sec.~\ref{sec:spin_waves}.
The effects of anisotropic interactions are discussed in Sec.~\ref{sec:anisotropy}.
Finally, Section~\ref{sec:conclusions} is devoted to concluding remarks and a summary of the results. 

\section{\label{sec:model_methods}Model and Methods}
\subsection{\label{subsec:model} Model}

The spin Hamiltonian for $\mathrm{Mn}_3\mathrm{X}$-type AB-stacked kagome lattice systems has been derived in Ref.~I from symmetry principles and contains four different types of interactions:

\begin{align}
    &\mathcal{H} = \mathcal{H}_{J} + \mathcal{H}_{D} + \mathcal{H}_{A} + \mathcal{H}_{K} \label{eq:magnetic_hamiltonian}\\
    &\mathcal{H}_{J} = \frac{1}{2}\sum_{\mathbf{r}\mathbf{r'}}\sum_{ij} J_{ij}(\mathbf{r}-\mathbf{r}') \mathbf{S}_{i}(\mathbf{r})\cdot\mathbf{S}_{j}(\mathbf{r'})\notag\\
    &\mathcal{H}_{D} = \frac{1}{2}\sum_{\mathbf{r}\mathbf{r'}}\sum_{ij} D_{ij}(\mathbf{r}-\mathbf{r}')\mathbf{\hat{z}}\cdot \left(\mathbf{S}_{i}(\mathbf{r})\times\mathbf{S}_{j}(\mathbf{r'})\right)\notag\\
    &\mathcal{H}_{A} = \frac{1}{2}\sum_{\mathbf{r}\mathbf{r'}}\sum_{ij}\sum_\alpha A_{ij\alpha}(\mathbf{r}-\mathbf{r'})\left(\mathbf{\hat{n}}_{i\alpha}\cdot\mathbf{S}_{i}(\mathbf{r})\right)\left(\mathbf{\hat{n}}_{j\alpha}\cdot\mathbf{S}_{j}(\mathbf{r'})\right),\notag\\
    &\mathcal{H}_{K} = \sum_\mathbf{r}\sum_{i}\sum_\alpha K_{\alpha} \left(\mathbf{\hat{n}}_{i\alpha}\cdot \mathbf{S}_{i}(\mathbf{r})\right)^2.\notag
\end{align} Here, $\mathcal{H}_{J}$ is the isotropic Heisenberg exchange, $\mathcal{H}_{D}$ is the DM interaction, $\mathcal{H}_{K}$ is the SIA, and $\mathcal{H}_{A}$ is the symmetric anisotropic exchange interaction.
Sum indices $\mathbf{r}$, $\mathbf{r'}$ label unit cells, $i, j\in\{1,...,6\}$ label atoms in each unit cell (Fig.~\ref{fig:structure} (c)), and $\alpha \in \{x,y,z\}$ labels the spin vector components.  
Vectors $\mathbf{n}_{i\alpha}$ represent local anisotropy axes and can be written as

\begin{equation}
    \mathbf{\hat{n}}_{ix} = \begin{bmatrix} \cos{\alpha_i} \\ \sin{\alpha_i} \\ 0\end{bmatrix},
    \mathbf{\hat{n}}_{iy} = \begin{bmatrix}-\sin{\alpha_i} \\ \cos{\alpha_i} \\ 0\end{bmatrix},
    \mathbf{\hat{n}}_{iz} = \begin{bmatrix}0 \\ 0 \\ 1\end{bmatrix},
\end{equation} 
where $\alpha_i$ give the angle of the anisotropy axes with respect to the global $x$-direction.
In this paper, we will restrict our attention to in- and out-of-plane nearest-neighbor (nn) and out-of-plane next-nearest-neighbor (nnn) interactions shown in Fig.~\ref{fig:structure} (a).
The interaction labels are based on $\mathrm{Mn}_3X$ bond distances, with index 1 labeling out-of-plane nn interactions, 2 and 3 labeling the in-plane nn, and 4 and 5 labeling the out-of-plane nnn.
In this work, we will ignore the breathing anisotropy~\cite{Chen_gs_2020_prb}, in order to simplify the analysis, in which case $\boldsymbol{\mathcal{A}}_2=\boldsymbol{\mathcal{A}}_3$, and $\boldsymbol{\mathcal{A}}_4=\boldsymbol{\mathcal{A}}_5$ (Fig.~\ref{fig:structure}).

It has recently been shown that an effective spin Hamiltonians of the form~(\ref{eq:magnetic_hamiltonian}) can be derived through perturbation theory from lattice Kondo model with SOC~\cite{Hayami_Yukitoshi_kondo_2018_prl,Yutaka_Ugadawa_Motome_kondo_2012_prl,Ghosh_kondo_2016_prb}.
The isotropic exchange terms in this case correspond to the Ruderman-Kittel-Kasuya-Yosida (RKKY) interactions while the remaining anisotropic spin interactions originate from the SOC.
However, while the DM interactions depend linearly on the strength of the SOC, the SIA and anisotropic exchange yield quadratic dependence~\cite{Moriya_dmi_1960_pr,Hayami_Yukitoshi_kondo_2018_prl}.
Therefore, within the perturbation theory, we generally expect the magnitudes of the latter two types of the anisotropic interactions to be smaller than that of the DM interaction.

It is sometimes useful to write the Hamiltonian in the general quadratic form:

\begin{equation}
    \mathcal{H} = \frac{1}{2}\sum_{\mathbf{r}\mathbf{r}'} \sum_{ij} \mathbf{S}^T_{i}(\mathbf{r}) \boldsymbol{\mathcal{A}}_{ij} (\mathbf{r}-\mathbf{r}') \mathbf{S}_{j}(\mathbf{r}'),
    \label{eq:hamiltonian_quadratic}
\end{equation} 
where the coupling matrix $\boldsymbol{\mathcal{A}}_{ij}(\boldsymbol{\delta})$ between two inequivalent spins is given by 

\begin{equation}
    \boldsymbol{\mathcal{A}}_{ij} (\boldsymbol{\delta}) = \begin{bmatrix}V_{ij}^{+}(\boldsymbol{\delta})& W_{ij}^{+}(\boldsymbol{\delta})& 0 \\ W_{ij}^{-}(\boldsymbol{\delta})& V_{ij}^{-}(\boldsymbol{\delta}) & 0 \\ 0 & 0 & X_{ij}(\boldsymbol{\delta})\end{bmatrix},
    \label{eq:general_coupling_matrix}
\end{equation} 
with

\begin{align}
    V_{ij}^{\pm}(\boldsymbol{\delta}) &= J_{ij}(\boldsymbol{\delta}) \pm A^{(xy)}_{ij}(\boldsymbol{\delta})\cos(\bar{\alpha}_{ij}),\notag\\
    W_{ij}^{\pm}(\boldsymbol{\delta}) &= \pm D_{ij}(\boldsymbol{\delta}) + A^{(xy)}_{ij}(\boldsymbol{\delta})\sin(\bar{\alpha}_{ij}),\notag\\
    X_{ij}(\boldsymbol{\delta}) &= J_{ij}(\boldsymbol{\delta}) + A^{(z)}_{ij}(\boldsymbol{\delta}).\notag
\end{align} 
Here, $\boldsymbol{\delta}=\mathbf{r}-\mathbf{r}'$, $\bar{\alpha}_{ij} = \alpha_i+\alpha_j$, $A_{ij}^{(xy)}$ and $A_{ij}^{(z)}$ are the two types of anisotropic exchange parameters allowed by symmetry.
The part of the coupling matrix that corresponds to the single-ion anisotropy is given by

\begin{equation}
    \boldsymbol{\mathcal{A}}_{ii} (0) = K^+\boldsymbol{\mathbb{I}} +  \begin{bmatrix}K^- \cos 2\alpha_i & K^- \sin 2\alpha_i  & 0 \\ K^- \sin 2\alpha_i  & -K^- \cos 2\alpha_i & 0 \\ 0 & 0 & K_Z\end{bmatrix},
    \label{eq:SIA_coupling_matrix}
\end{equation} 
where $\boldsymbol{\mathbb{I}}$ is a $3\times 3$ identity matrix, $K^+ = K_x+K_y$, $K^- = K_x-K_y$, and $K_Z = 2K_z - K^+$.
Note that the first term in this expression is just a constant energy shift, and therefore $K^+$ can be ignored in further calculations.

\subsection{\label{subsec:methods_LT}Luttinger-Tisza method}

To provide an initial characterization of the classical ground states, we define lattice Fourier transforms of the spin vectors as

\begin{equation}
    \mathbf{S}_i(\mathbf{r}) = \frac{1}{\sqrt{N}}\sum_\mathbf{q} \mathbf{S}_i(\mathbf{q}) e^{-i\mathbf{q}\cdot\mathbf{r}},
    \label{eq:spin_FT}
\end{equation} 
where $\mathbf{q}$ are the wave-vectors restricted to the first Brillouin zone, and $\mathbf{S}(\mathbf{q})$ are the Fourier amplitudes.
From Eq.~(\ref{eq:magnetic_hamiltonian}) and (\ref{eq:hamiltonian_quadratic}), the total energy of the system is

\begin{equation}
    \mathcal{H} = \frac{1}{2}\sum_{\mathbf{q}} \sum_{ij} \mathbf{S}^T_{i}(\mathbf{q}) \boldsymbol{\mathcal{A}}_{ij} (\mathbf{q}) \mathbf{S}_{j}(-\mathbf{q}),
    \label{eq:hamiltonian_q}
\end{equation} 
where the Fourier transform of the magnetic interactions are given by 

\begin{equation}
    \boldsymbol{\mathcal{A}}_{ij} (\mathbf{q}) = \sum_{\boldsymbol{\delta}} \boldsymbol{\mathcal{A}}_{ij} (\boldsymbol{\delta}) e^{-i\mathbf{q}\cdot\boldsymbol{\delta}}.
\end{equation} 
The true ground state is calculated by minimizing ~(\ref{eq:hamiltonian_q}), subject to local normalization constraints $|\mathbf{S}_i(\mathbf{r})| = 1$ for all spins in the system.
This strong constraint significantly complicates the problem and often makes it impossible to solve.
Instead, the Luttinger-Tisza (LT) method~\cite{Luttinger_Tisza_1946_pr} replaces the local constraints by a global constraint, whereby the sum of all spin magnitudes is set to be equal to the number of spins.
This simplification allows one to recast the energy minimization in the form of an eigenvalue problem,

\begin{equation}
    \sum_{j} \boldsymbol{\mathcal{A}}_{ij} (\mathbf{q}) \mathbf{S}_{j}(\mathbf{q}) = \varepsilon(\mathbf{q})\mathbf{S}_{i}(\mathbf{q})
\end{equation} 
Under ideal circumstances, the smallest eigenvalue $\varepsilon_\text{LT}(\mathbf{q})$ and the corresponding eigenvector give the ground state of the magnetic system.
However, this method often produces unphysical solutions for systems with strong anisotropic interactions and multiple sublattices~\cite{Zaliznyak_Zhitomirsky_LT_2003_arxiv,Maximov_Chernyshev_2019_prx}. 
As a result, we used the LT method to determine the approximate locations of the phase boundaries, as well as the lower bounds on the ground state energies to guide the numerical calculations. 

\subsection{\label{subsec:methods_MC}Monte Carlo}

To identify the ground state magnetic configurations, we utilized classical Monte Carlo (MC) simulations, which were carried out using standard local heat-bath updates.
To resolve the individual phases, we have used system sizes ranging from $6^3$ to $24^3$ unit cells (1296 to 82944 spins respectively) and between $10^4$ and $10^6$ MC steps.
In each simulation, the temperature is reduced down to $T\approx 10^{-6}$, to ensure energy convergence.
All phases presented in this work have $Q_z=0$, meaning that every unit cell along the $c$-axis has exactly the same magnetic structure.
This fact allowed us to determine the phase boundaries using smaller system sizes (between $6\times 6\times 2$ and $18\times 18\times 2$ unit cells). 
The MC data is Fourier transformed to obtain the spin structure factor

\begin{equation}
    S(\mathbf{q}) = \frac{1}{6}\sum_{ij} \big\langle \mathbf{S}_{i}(\mathbf{q})\cdot\mathbf{S}_{j}(-\mathbf{q})\big\rangle e^{i\mathbf{q}\cdot(\mathbf{r}_i-\mathbf{r}_j)},
\end{equation} 
where $\mathbf{r}_i$ are the positions of atoms inside of the unit cell.

\subsection{\label{subsec:methods_dynamics}Spin waves}

The dynamics of the magnetic system can be described by the Landau-Lifshitz equation

\begin{equation}
    \frac{\text{d}\mathbf{S}_i(\mathbf{r},t)}{\text{d}t} = \mathbf{H}_{i}(\mathbf{r},t)\times \mathbf{S}_i(\mathbf{r},t),
    \label{eq:LL_equation}
\end{equation} 
where $\mathbf{H}_{i}(\mathbf{r})$ is the effective field at each site

\begin{equation}
    \mathbf{H}_{i}(\mathbf{r},t) = \sum_{\boldsymbol{\delta}} \sum_j \boldsymbol{\mathcal{A}}_{ij}(\boldsymbol{\delta}) \mathbf{S}_{j}(\mathbf{r}+\boldsymbol{\delta},t).
\end{equation} 
In this work, we will look for solutions of the linearized form of Eq.~(\ref{eq:LL_equation}) which correspond to the low-energy spin wave excitations.
This is done by first changing into a local coordinate system where the local $z$-components are aligned with the ground state spin configuration.
Note that as long as this coordinate transformation is described by a local rotation $\mathbf{U}_i(\mathbf{r})$, the dynamic evolution of the local spin components can be written in the same form as Eq.~(\ref{eq:LL_equation}), replacing $\mathbf{S}_i(\mathbf{r},t)$ and $\mathbf{H}_i(\mathbf{r},t)$ with $\widetilde{\mathbf{S}}_i(\mathbf{r},t) = \mathbf{U}_i(\mathbf{r})\mathbf{S}_i(\mathbf{r},t)$ and $\widetilde{\mathbf{H}}_i(\mathbf{r},t) = \mathbf{U}_i(\mathbf{r})\mathbf{H}_i(\mathbf{r},t)$ respectively.

\section{\label{sec:symmetry}Spin symmetry}

When studying the properties of a magnetic model, it is important to take a proper account of the symmetries that leave the Hamiltonian invariant.
Apart from the space group transformations imposed by the underlying crystalline lattice, magnetic systems also include symmetries associated with spin rotations and reflections.
The most common spin symmetry is the time reversal, $T$, which must be broken in order to establish magnetic order.
A combination of crystallographic group operations with the time reversal operator leads to \textit{magnetic point and space groups}.
The addition of this one operator significantly extends the number of symmetrically distinct systems: in three dimensions there are 1651 magnetic space groups compared to 230 ``regular'' crystal space groups~\cite{Glazer_Burns_space_groups_book_2013,bradley_cracknell_space_groups_book_2010}.
In materials with large SOC, the spins are typically pinned to the lattice, meaning that for each lattice transformation there is a corresponding spin transformation.
The symmetry group of the Hamiltonian is then the \textit{paramagnetic group} which is a direct product 

\begin{equation}
    \mathcal{G}_\text{SOC} = \mathcal{G}_L\otimes Z_2^{(T)},
\end{equation}
where $ \mathcal{G}_L$ is the space group and $Z_2^{(T)} = \{E,T\}$.
However, in the limit of decoupled spin and orbital degrees of freedom one has a magnetic system with isotropic Heisenberg exchange interactions which are invariant under all global spin rotations. 
In fact, as was noted by Brinkman and Elliott, there are many instances of magnetic systems where the spins are at least partially decoupled from the lattice~\cite{Brinkman_Elliot_Roger_Peierls_spin_groups_1966_prsl,Brinkman_Elliot_spin_groups_1966_jap,Brinkman_spin_groups_1967_jap}.
The symmetry of these systems is then described by the \textit{spin point and space groups}~\cite{Litvin_Opechowski_spin_groups_1974_physica,Litvin_spin_groups_1977_acsa} which are typically much larger than the corresponding magnetic groups.
Although the effects of the extended spin symmetry operations have previously been considered almost exclusively in the context of either isotropic spins or single-ion anisotropy, their importance in the intermediate cases, which include DM and anisotropic exchange interactions, remains relatively underrepresented and has only began gaining interest in recent years~\cite{Corticelli_Moessner_McClarty_spin_groups_2022_prb,Liu_Liu_spin_groups_2022_prx}.
For more information about the spin space groups we refer the reader to reference~\cite{Corticelli_Moessner_McClarty_spin_groups_2022_prb} which provides an excellent review of the subject. 
In compounds with $\mathrm{Mn}_3X$-like structure, depending on the strength of the SOC, one can identify three distinct cases for magnetic models depicted in Fig.~\ref{fig:SOC_symmetry_diagram}. 
Each case corresponds to a different group of spin symmetries.


In this section, we present the symmetry analysis of these three cases by deriving the corresponding spin groups and determining the resulting irreps for $\mathbf{Q}=0$.
Note that the analysis presented here is for classical magnetic moments (rotations in SO(3)), but could be readily extended to quantum spin operators (rotations in SU(2)).
The details of some derivations can be found in the [\textbf{Supplemental Material}].

\subsection{\label{subsec:symmetry_decoupled}Decoupled case}

\begin{figure*}[!t]
    \centering
    \includegraphics[width=0.95\textwidth]{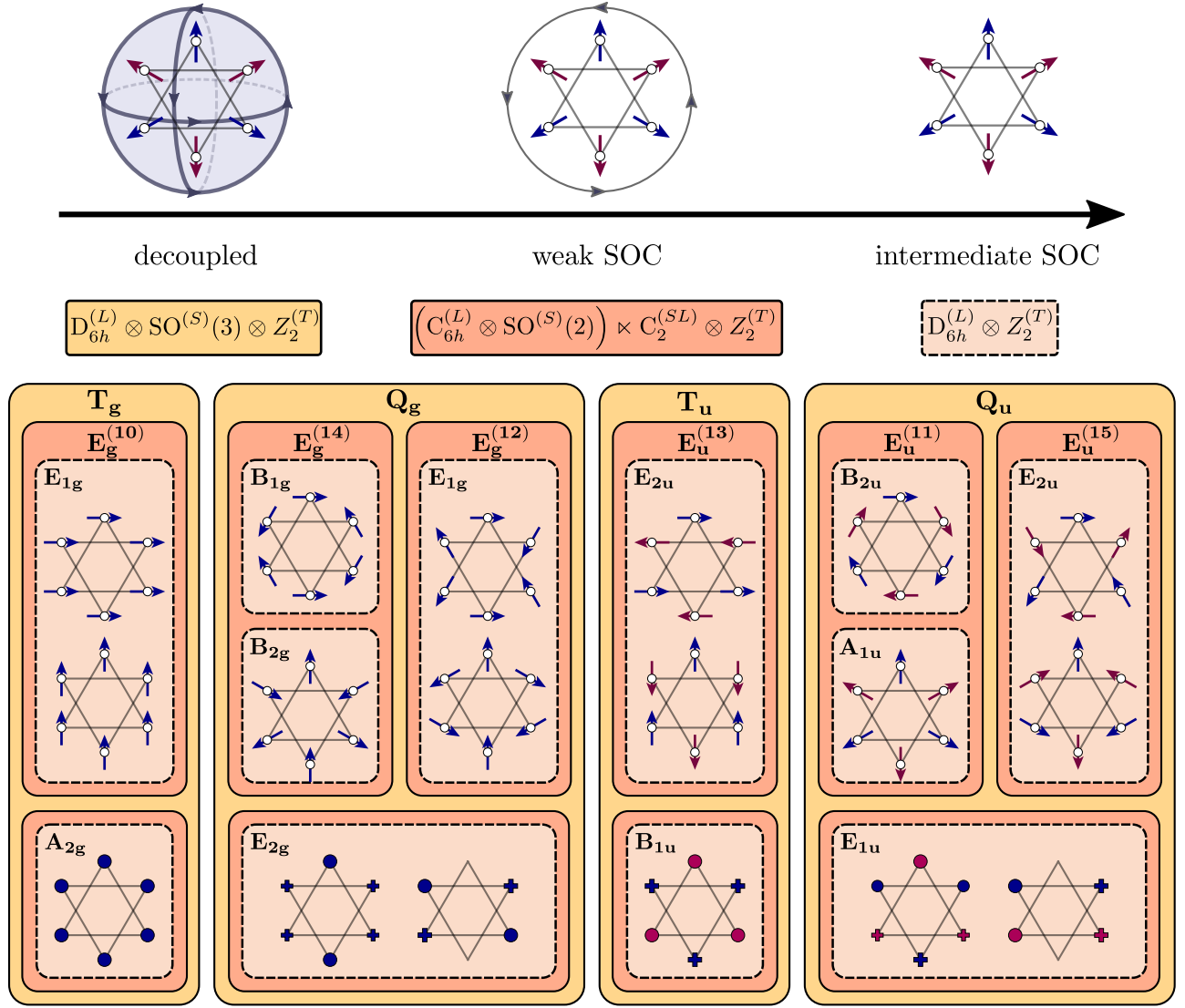}
    \caption{A diagram illustrating the effects of SOC on the symmetry of the spin Hamiltonian. In $\mathrm{Mn}_3X$ systems, we identify three SOC limits resulting in distinct symmetry groups: decoupled, weak SOC, and intermediate SOC. These correspond to, in the same order, the isotropic, XY-anisotropy, and Ising anisotropy (top panel). The total spin group of the Hamiltonian and the corresponding irreps are indicated by the colored blocks. The positive and negative $z$-components of spins are indicated by the filled circles and pluses respectively, and the blue and red colors of the spin components are used to indicate parallel and antiparallel out-of-plane nnn (labels ``g'' and ``u'' respectively). There are four irreps in the decoupled limit, which are first split into ten irreps in the weak SOC limit and then further split into twelve irreps in the intermediate SOC limit. The irreps with the out-of-plane spin components (bottom row) are unchanged going from the weak to intermediate SOC limit, and therefore have the same labels. The remaining planar irreps are labeled using two labels corresponding to the weak and intermediate limits.}
    \label{fig:SOC_symmetry_diagram}
\end{figure*}

As mentioned previously, when the spin and orbital degrees of freedom are completely decoupled, the magnetic interactions correspond to the isotropic exchange, $\mathcal{H} = \mathcal{H}_J$, meaning that the spin Hamiltonian is invariant with respect to pure space transformations (lattice site permutations) and global spin rotations by an arbitrary angle.
The crystal symmetries pertaining to this work form a space group P6$_3/mmc$, which we denote as $\mathrm{D}_{6h}^{(L)}$ (indicating also the point group), and the spin rotations combined with the time reversal symmetry form a group $\text{SO}^{(S)}(3)\otimes Z_2^{(T)}$.
Since the elements in these two groups commute, the full group of the isotropic exchange Hamiltonian is simply 
\begin{equation}
    \mathcal{G}_J = \mathrm{D}_{6h}^{(L)}\otimes \mathrm{SO}^{(S)}(3)\otimes Z_2^{(T)}.
\end{equation} 
As a result, the irreps of $\mathcal{G}_J$ are obtained by taking a direct product of the irreps of spin and lattice degrees of freedom.
The former is related to the set of spherical harmonics with $l=1$ and has dimension 3, while the latter depends on the periodicity of the magnetic structure.
For $\mathbf{Q}=0$, the symmetries of $\mathrm{D}_{6h}^{(L)}$ reduce to the point group, and the irrep decomposition becomes $A_{1g}\oplus E_{2g}\oplus B_{2u}\oplus E_{1u}$.
Therefore, there are four irreps in the decomposition of magnetic states: two of dimension 3 ($T_g$, $T_u$), and two of dimension 6 ($Q_g$, $Q_u$), as illustrated in Fig.~\ref{fig:SOC_symmetry_diagram}.
The two triplets correspond to the collinear configurations, while the two six-dimensional irreps are related to the 120 degree configurations. 
The labels ``g'' and ``u'' are used to indicate the parity of the irreps under the spatial inversion, as per usual group theory notation.

\begin{table*}[t]
    \centering
    \resizebox{0.8\textwidth}{!}{\begin{tabular}{c|r r r r r r r r r r r r}
        $\mathcal{G}_D$ & $C_1(\phi)$ & $C_6(\phi)$ & $C_3(\phi)$ & $C_2(\phi)$ & $C_2'$ & $C_2''$ & $I(\phi)$ & $IC_6(\phi)$ & $IC_3(\phi)$ & $IC_2(\phi)$ & $IC_2'$ & $IC_2''$ \\
        \hline
        $E_g^{(nm)}$ & $c^{(0)}_n$ & $c^{(m)}_n$ & $c^{(2m)}_n$ & $c^{(3m)}_n$ & 0 & 0 & $c^{(0)}_n$ & $c^{(m)}_n$ & $c^{(2m)}_n$ & $c^{(3m)}_n$ & 0 & 0 \\
        $E_u^{(nm)}$ & $c^{(0)}_n$ & $c^{(m)}_n$ & $c^{(2m)}_n$ & $c^{(3m)}_n$ & 0 & 0 &-$c^{(0)}_n$ &-$c^{(m)}_n$ & -$c^{(2m)}_n$ & -$c^{(3m)}_n$ & 0 & 0
    \end{tabular}}
    \caption{Characters of the two-dimensional irreps of group $\mathcal{G}_D$ that do not also appear in group $\mathrm{D}_{6h}\otimes Z_2^{(T)}$. For simplicity, the elements with time reversal operator are not included in the table. Here, the classes $C_k(\phi)$ have two elements, $\{C_z(\pm \phi,\pm \phi_k)\}$, which corresponds to a rotation around $z$-axis of spin by an angle $\phi$ and lattice by angle $\phi_k = \frac{2\pi}{k}$. $C_2'$ and $C_2''$ include simultaneous 180 degree rotations of lattice (same as in the $\mathrm{D}_{6h}$ point group) and spin with axes in the $xy$-plane. The second half of classes is obtained by combining the first half with the spatial inversion operator. The non-zero characters are defined as $c_n^{(m)} = 2\cos\left(n\phi+\frac{m\pi}{3}\right)$, where $n \in \mathbb{Z}\setminus\{0\}$, $m \in \mathbb{Z}_6$.}
    \label{tab:GD_character_table}
\end{table*}

\subsection{\label{subsec:symmetry_weakly_coupled}Weak coupling case}

As discussed in Sec.~\ref{sec:model_methods}, when the SOC is small but non-zero, it is reasonable to assume that the DM interactions are the dominant type of anisotropy in the system, giving $\mathcal{H}\approx\mathcal{H}_J + \mathcal{H}_D$.
The addition of DM coupling to the spin Hamiltonian significantly complicates the symmetry analysis of the model.
However, it is still possible to determine the structure of the corresponding spin group as well as all irreps.
We also note that the SIA and bond-dependent interactions that couple the $z$-components of spins do not change the symmetry of this spin group.  

It can be shown that a spin rotation applied to a DM coupling between two spins gives

\begin{align}
    D\mathbf{\hat{z}}\cdot[\mathbf{S}_i'(\mathbf{r})\times\mathbf{S}_j'(\mathbf{r}')] &= D\mathbf{\hat{z}}\cdot[(\mathbf{M}\mathbf{S}_i(\mathbf{r}))\times(\mathbf{M}\mathbf{S}_j(\mathbf{r}'))]\notag\\
    &= D(\mathbf{M}^T\mathbf{\hat{z}})\cdot[\mathbf{S}_i(\mathbf{r})\times\mathbf{S}_j(\mathbf{r}')],
\end{align} 
where $\mathbf{M}$ is the rotation matrix.
Therefore, all rotations that leave $\mathbf{\hat{z}}$ invariant belong to the symmetry group of DMI.
This constitutes a group of axial rotations in spin space $\mathrm{SO}^{(S)}(2)$, implying XY-anisotropy.
The complications arise from the fact that not all lattice permutations leave the DM interaction invariant. 
In particular, transformations that include $C_2$ rotations around axes parallel to the kagome layers, and the corresponding reflections reverse the direction of the bonds, which flips the sign of the DM vector.
In order for these to become proper symmetry operators, they must be combined with the corresponding spin rotations/reflections.
Since the spin and lattice operations are now coupled and do not necessarily commute with each other, we can no longer write the total group of the Hamiltonian as a direct product of spin and lattice symmetry groups.
Nevertheless, the group of DM coupling can then be written as a \textit{semidirect} product

\begin{equation}
    \mathcal{G}_D = \left(\mathrm{C}_{6h}^{(L)}\otimes \mathrm{SO}^{(S)}(2)\right)\ltimes \mathrm{C}_2^{SL}\otimes Z_2^{(T)},
\end{equation} 
where $\mathrm{C}_{6h}^{(L)}$ is the group of lattice symmetries that leave DM interaction invariant, and $\mathrm{C}_2^{(SL)}$ has one non-trivial element $C_2^S C_2^L$ that rotates both the lattice and spin components around the $x$-axis.
The derivation of the group structure and irreps of $\mathcal{G}_D$ is given in the [\textbf{Supplemental Material}].
The irreps of $\mathcal{G}_D$ include all irreps of a regular $\mathrm{D}_{6h}\otimes Z_2^{(T)}$ magnetic group, and an infinite number of two-dimensional irreps, as presented in table~\ref{tab:GD_character_table}.

Intuitively, one can expect that the continuous axial rotational symmetry would separate the $z$-components of the spins from the planar components, while leaving the latter degenerate.
The decomposition of a $\mathbf{Q}=0$ magnetic structure consists of ten irreps, four of which involve only the $z$-components of spins ($A_{2g}\oplus E_{2g}\oplus B_{1u}\oplus E_{1u}$), and six two-dimensional planar irreps ($E_g^{(10)}\oplus E_g^{(12)}\oplus E_g^{(14)}\oplus E_u^{(11)}\oplus E_u^{(13)}\oplus E_u^{(15)}$), as shown in Fig.~\ref{fig:SOC_symmetry_diagram}.
Note that the order parameters corresponding to the two-dimensional out-of-plane irreps ($E_{2g}$ and $E_{1u}$) do not have fixed norms and therefore by themselves cannot be observed in a classical system~\cite{Essafi_Benton_Jaubert_duality_2017_prb}.
The labels $E_a^{(nm)}$ provide information about the parity of the magnetic order parameters (label $a$), transformation properties under spin rotations (subscript $n$), and the coupling between the spin and spatial transformations (subscript $m$).
In the case of the planar irreps, the $m$ subscript can also be viewed as the ``winding'' of the spins around the hexagon.

\subsection{\label{subsec:symmetry_strongly_coupled}Intermediate coupling case}

Finally, when the SOC is sufficiently strong, the SIA and bond-dependent anisotropies become important, effectively reducing the symmetry of the spin Hamiltonian to the magnetic group $\mathrm{D}_{6h}\otimes Z_2^{(T)}$.
As noted before, only those anisotropic interactions involving the planar spin components explicitly break the global axial rotation symmetry, forcing the spins to align with the local anisotropy axes $\mathbf{\hat{n}}_{i\alpha}$.
Therefore, these interactions are inherently Ising-like.
This anisotropy splits the $E_g^{(14)}$ and $E_u^{(11)}$ configurations each into two singlets, while leaving the degeneracy of the remaining irreps unchanged (Fig.~\ref{fig:SOC_symmetry_diagram}). 

\section{\label{sec:self_duality}Self-duality transformations}

In order to systematically describe the classical ground state properties of a magnetic model, one must address the problem of the dimensionality of the parameter space.
In the general case considered in this work, there are more than a dozen independent parameters, making the complete computational analysis of the phase diagram forbiddingly expensive.
As a result, it is necessary to determine the means of reducing the parameter space.
The most common approach is to focus on a particular physical example (\textit{e.g.} a family of compounds) where the ranges of the coupling constants are approximately known from either the experimental data or from \textit{ab initio} calculations. 
This approach allows one to bound the values of the parameters, and potentially even ignore some of them.
Although this method is of extreme utility for explaining the properties of the known compounds, it may provide very limited information for describing the functionality of novel compounds, since the relevant parts of the parameter space may fall far beyond the explored subspace. 
Moreover, even after the relevant parameter ranges are identified, the dimensionality of the search space may be quite large and still require extensive computations.
For example, in the weak SOC limit, the magnetic model considered in this work still contains 4-5 independent interactions.

Another approach, often neglected in the literature, is to determine hidden relationships between the models with different coupling constants.
It is often true that the parameters in a given model are not completely independent, and one can determine a set of transformations that map the Hamiltonian onto itself, while changing the values of the coupling constants.
Such transformations are referred to as the \textit{self-duality} transformations and are the subject of this section.
We note that while the two methods described here are different in nature, one can and should use them together to achieve a systematic yet physically relevant description of a spin model. 
In this work, we constrain the values of the coupling constants, in particular those originating from the SOC, by referring to the experimental and numerical results for the $\mathrm{Mn}_3X$ compounds~\cite{Soh_gs_2020_prb,Chen_gs_2020_prb,Park_gs_2018_nature_pub}. 

Self-duality transformations have played an important role in statistical physics, an important example being a Kramers-Wannier duality that relates the ordered and paramagnetic phases in the two-dimensional Ising model on a square lattice~\cite{Kramers_Wannier_duality_1941_pr,Onsager_ising_1944_pr,Wannier_duality_1945_rmp,Wegner_duality_ising_1971_jmp,Wu_Wang_duality_1976_jmp,Savit_duality_1980_rmp}. 
Self-duality maps provide a natural formulation of a renormalization flow and have therefore been used in studies of critical phenomena. 
More recently, a different class of self-dual transformations has been derived for Heisenberg-Kitaev models on honeycomb and triangular lattices~\cite{Chaloupka_Khaliullin_duality_honeycomb_2015_prb,Kimchi_Vishwanath_duality_lattices,Maximov_Chernyshev_2019_prx}. 
These transformations have been referred to in the literature as the Klein duality, since they form a group isomorphic to the Klein group.
The main interest in the self-duality maps has been the search of accidental degeneracy points, where strongly anisotropic systems at times display full rotational symmetry~\cite{Chaloupka_Khaliullin_duality_honeycomb_2015_prb,Maximov_Chernyshev_2019_prx}.
Self-duality in two-dimensional kagome layers has been considered before in~\cite{Essafi_Benton_Jaubert_duality_2016_nature_comm,Essafi_Benton_Jaubert_duality_2017_prb}.
There, it was used to draw connections between different models that support spin liquid phases.

In this section, we determine the relevant self-duality transformations for Hamiltonian~(\ref{eq:magnetic_hamiltonian}).
Since the number of possible self-dualities depends on the types of magnetic interactions in the model, we separate the discussion into three parts corresponding to the three types of spin Hamiltonians discussed in Sec.~\ref{sec:symmetry} (see Fig.~\ref{fig:SOC_symmetry_diagram}).

\subsection{\label{subsec:duality_derivation} Self-duality as a permutation of spin invariants}

For the purposes of this work, we define a self-duality transformation as a simultaneous transformation of the spin variables and model parameters that leaves the Hamiltonian unchanged:

\begin{equation}
    \begin{split}
        \mu: \{\mathbf{S}_{i}(\mathbf{r});\boldsymbol{\mathcal{A}}_{ij}(\boldsymbol{\delta})\} \longrightarrow \{\widetilde{\mathbf{S}}_{i};\boldsymbol{\widetilde{\mathcal{A}}}_{ij}(\boldsymbol{\delta})\}, \\
        \mathcal{H}\big(\{\mathbf{S}_{i};\boldsymbol{\mathcal{A}}_{ij}(\boldsymbol{\delta})\}\big) = \mathcal{H}\big(\{\widetilde{\mathbf{S}}_{i};\boldsymbol{\widetilde{\mathcal{A}}}_{ij}(\boldsymbol{\delta})\}\big).
    \end{split}
    \label{eq:duality_definition}
\end{equation}
The transformation in Eq.~(\ref{eq:duality_definition}) is implied to include all spins in the system, as well as all symmetry-allowed coupling constants.
We assume that the spin transformations can be expressed as a site-dependent linear operation $\widetilde{\mathbf{S}}_i(\mathbf{r}) = \mathbf{M}_i^T(\mathbf{r})\mathbf{S}_i(\mathbf{r})$, where $\mathbf{M}_i(\mathbf{r})$ is an orthogonal matrix.
Then, 
\begin{equation}
    \boldsymbol{\widetilde{\mathcal{A}}}_{ij}(\mathbf{r}-\mathbf{r}') = \mathbf{M}_i^T(\mathbf{r})\boldsymbol{\mathcal{A}}_{ij}(\mathbf{r}-\mathbf{r}') \mathbf{M}_j(\mathbf{r}').
\end{equation}
We require that the matrices $\boldsymbol{\mathcal{A}}_{ij}(\boldsymbol{\delta})$ and $\boldsymbol{\widetilde{\mathcal{A}}}_{ij}(\boldsymbol{\delta})$ only differ in the values of the coupling parameters. 
Note that matrices $\mathbf{M}_i(\mathbf{r})$ are not necessarily unique: we can define $\widetilde{\mathbf{M}}_i(\mathbf{r}) = \mathbf{R}\mathbf{M}_i(\mathbf{r})$ where $\mathbf{R}$ is a symmetry operation that leaves $\boldsymbol{\mathcal{A}}_{ij}(\boldsymbol{\delta})$ invariant.

To demonstrate the relationship between the symmetry of the system and the number of self-duality transformations, we first note that any quadratic spin Hamiltonian of the form (\ref{eq:hamiltonian_quadratic}) can be written as a sum of bilinear spin invariants:

\begin{equation}
    \mathcal{H} = \sum_\lambda \mathcal{A}^{(\lambda)} B^{(\lambda)},
\end{equation}
where $B^{(\lambda)}$ are the invariants, and $\mathcal{A}^{(\lambda)}$ are the corresponding coupling constants.
One can show~[\textbf{Supplemental Material}] that the bilinear spin invariants can be calculated by squaring the symmetry-adapted order parameters corresponding to the irreps in the decomposition of the spin structure (see Sec.~\ref{sec:symmetry}).
The result can be written as

\begin{equation}
    B^{(\lambda)} = B_{kl}^{(\Gamma)} = \mathbf{S}_k^{(\Gamma)}\cdot\mathbf{S}_l^{(\Gamma)},
\end{equation}
where $\mathbf{S}_k^{(\Gamma)}$ is the symmetry-adapted order parameter corresponding to irrep $\Gamma$, and $k, l$ label different order parameters belonging to $\Gamma$.
The number of components in this vector is equal to the dimensionality of the $\Gamma$.

It can be shown that a transformation of the spins that results in a permutation of the invariants,

\begin{equation}
    \mu\left(B^{(\lambda)}\right) = \sum_\lambda P_{\lambda\mu(\lambda)}B^{(\lambda)} = B^{(\mu(\lambda))},
\end{equation}
satisfies our definition of self-duality since

\begin{equation}
    \mu\left(\mathcal{H}\right) = \sum_\lambda \mathcal{A}^{(\lambda)} B^{(\mu(\lambda))} = \sum_\lambda \widetilde{\mathcal{A}}^{(\lambda)} B^{(\lambda)}.
\end{equation}
Here, $P_{\lambda\mu(\lambda)}$ permutes indices $\lambda$ and $\mu(\lambda)$, and 

\begin{equation}
    \widetilde{\mathcal{A}}^{(\lambda)} = \mu^{-1}\left(\mathcal{A}^{(\lambda)}\right).
\end{equation}
At the same time, a permutation of invariants occurs when we permute the order parameters, and possibly change the sign of some of the components:

\begin{equation}
    \mu:S_{k\alpha}^{(\Gamma)} \longrightarrow \pm S_{k'\alpha}^{(\Gamma')},
\end{equation}
where $\alpha$ labels the components of the order parameters,
Therefore, we look for the spin transformations $M_i(\mathbf{r})$ that correspond to permutations between order parameters.  
Although this is by no means a rigorous derivation, the self-dualities determined this way are sufficient to describe most interesting properties observed in this paper \footnote{We note that up to now the discussion has been completely general and is applicable to any spin system described by a set of symmetries. The derivation of the general conditions for self-duality go beyond the scope of this work and are a subject of an ongoing investigation.}.
In the remainder of this section we present the relevant duality transformations in the context of the three SOC cases discussed in the last section.

As a final note, we point out that the set of all self-duality maps resulting from the permutations of the symmetry-adapted order parameters forms a group.
This can be deduced from the fact that we only allow orthogonal matrices as the local transformations of spins.
This fact significantly simplifies our search, since the new transformations can be obtained by combining together those already identified in the analysis.

\subsection{\label{subsec:duality_J}Decoupled case}


In the decoupled limit, the coupling matrices are diagonal in the spin components:

\begin{equation}
    \boldsymbol{\mathcal{A}}_{ij}(\boldsymbol{\delta}) = J_{ij}(\boldsymbol{\delta}) \boldsymbol{\mathbb{I}},
\end{equation}
where $\boldsymbol{\mathbb{I}}$ is an identity matrix.
As a result, there is a single self-duality transformation, which corresponds to transformations between order parameters symmetric and antisymmetric under inversion operation:

\begin{equation}
    \gamma^{(-1)} : \begin{cases} T_g \longleftrightarrow T_u,\\ Q_g \longleftrightarrow Q_u. \end{cases}
\end{equation}
The corresponding group of spin transformations can be written as:

\begin{equation}
    \mathbf{M}_i^{(g)} = \begin{cases}g\boldsymbol{\mathbb{I}}& \text{if } i\in\{1,2,3\},\\ \phantom{-}\boldsymbol{\mathbb{I}}& \text{if } i\in\{4,5,6\},\end{cases}
\end{equation}
where $g=\pm 1$.
$\mathbf{M}_i^{(+1)}(\mathbf{r})$ is the identity, and $\mathbf{M}_i^{(-1)}(\mathbf{r})$ flips all spins in layer A, while keeping the layer B unchanged.
While this transformation does not affect the in-plane interactions, the out-of-plane couplings change sign
\begin{equation}
    \gamma^{(-1)} : \begin{cases} J_1 \longrightarrow -J_1,\\ J_2 \longrightarrow \phantom{-}J_2,\\ J_4 \longrightarrow -J_4. \end{cases}
\end{equation}
Thus, $\gamma^{(-1)}$ provides a map between models with ferro- and antiferromagnetic out-of-plane interactions.

\subsection{\label{subsec:duality_JD}Weak coupling case}

Next, consider a model with weak SOC.
The coupling matrix between two spins~(\ref{eq:general_coupling_matrix}) can then be simply written as 

\begin{equation}
    \boldsymbol{\mathcal{A}}_{ij}(\boldsymbol{\delta}) = \begin{bmatrix} J_{ij}(\boldsymbol{\delta}) & D_{ij}(\boldsymbol{\delta}) & 0\\
   -D_{ij}(\boldsymbol{\delta}) & J_{ij}(\boldsymbol{\delta}) & 0\\
    0 & 0 & J_{ij}(\boldsymbol{\delta}) + A_{ij}^{(z)}(\boldsymbol{\delta})\end{bmatrix}.
\end{equation}
We must formally include $A_{ij}^{(z)}(\boldsymbol{\delta})$ anisotropic interactions in some cases in order to define proper self-duality transformations.
However, as will become clear in the next sections, this formality is not too significant in practice, since the majority of the observed phases in this work are restricted to the $xy$-plane.
Therefore, many useful properties of self-duality relating to the spin structures and phase diagrams remain exact even if we ignore $A_{ij}^{(z)}(\boldsymbol{\delta})$, since the $z$-components of the spins are decoupled from the planar components (Fig.~\ref{fig:SOC_symmetry_diagram}), and so the duality transformations are also decoupled.

We first focus on the permutations of the four out-of-plane order parameters.
There is a single non-trivial duality transformation corresponding to flipping the $z$-components in the A layer.
The group of transformations is then written as

\begin{equation}
    \mathbf{M}_i^{(\eta)} = \begin{bmatrix} 1 & 0 & 0\\ 0 & 1 & 0\\ 0 & 0 & \eta_i \end{bmatrix},
\end{equation}
where

\begin{equation}
    \eta_i = \begin{cases}\eta & \text{if } i\in\{1,2,3\},\\ 1& \text{if } i\in\{4,5,6\},\end{cases}
\end{equation}
and $\eta = \pm 1$.
The parameters are mapped according to

\begin{equation}
    \mu^{(-1)} : \begin{cases} J_i &\longrightarrow \phantom{-}J_i,\\ A_1^{(z)} &\longrightarrow -A_1^{(z)}-2J_1,\\ A_2^{(z)} &\longrightarrow \phantom{-}A_2^{(z)},\\ A_4^{(z)} &\longrightarrow -A_4^{(z)}-2J_4. \end{cases}
\end{equation}

Next, we identify the transformations that permute the six planar order parameters. 
As discussed in the previous section, these are labeled by an integer $m$ which represents the six-fold winding number.
Therefore, the first natural choice for a spin transformation is a local rotation written as

\begin{equation}
    \mathbf{M}^{(m)}_i = \begin{bmatrix} \cos\theta_i^{(m)} &-\sin\theta_i^{(m)} & 0\\ 
    \sin\theta_i^{(m)} & \phantom{-}\cos\theta_i^{(m)} & 0\\
    0 & 0 & 1\end{bmatrix},
    \label{eq:matrix_dual_m}
\end{equation}
where

\begin{equation}
    \theta_i^{(m)} = \frac{\pi ml_i}{3},
    \label{eq:duality_angles}
\end{equation} 
$m\in\mathbb{Z}_6$ and $l_i$ label the positions of atoms on the hexagon of the unit cell in the counter clock-wise direction:

\begin{equation}
    \{l_1,l_2,l_3,l_4,l_5,l_6\} = \{3,1,5,0,4,2\}.
\end{equation} 
However, there is another group of global transformations that leads to a distinct duality:

\begin{equation}
    \mathbf{M}^{(\varepsilon)} = \begin{bmatrix} \varepsilon & 0 & 0\\ 
    0 & 1 & 0\\
    0 & 0 & 1\end{bmatrix},
\end{equation}
with $\varepsilon=\pm 1$.
Note that the elements of $\mathbf{M}_i^{(m)}$ and $\mathbf{M}^{(\varepsilon)}$ in general do not commute.
We use elements $\mathbf{M}^{(\varepsilon)}\mathbf{M}_i^{(m)}$ to define the self-duality transformations $\mu_m^{(\varepsilon)}$, which transform the coupling constants according to

\begin{equation}
    \mu_m^{(\varepsilon)}:
    \begin{dcases}
        J_1 \longrightarrow J_1\cos{\left(\frac{\pi m}{3}\right)} - D_1\varepsilon\sin{\left(\frac{\pi m}{3}\right)},\\
        D_1 \longrightarrow J_1\sin{\left(\frac{\pi m}{3}\right)} + D_1\varepsilon\cos{\left(\frac{\pi m}{3}\right)},\\
        J_2 \longrightarrow J_2\cos{\left(\frac{2\pi m}{3}\right)} - D_2\varepsilon\sin{\left(\frac{2\pi m}{3}\right)},\\
        D_2 \longrightarrow J_2\sin{\left(\frac{2\pi m}{3}\right)} + D_2\varepsilon\cos{\left(\frac{2\pi m}{3}\right)},\\
        J_4 \longrightarrow J_4(-1)^m.
    \end{dcases}
    \label{eq:duality_weak_SOC}
\end{equation}
In addition, if $\widetilde{J}_i$ is the new exchange coupling constant, then

\begin{equation}
    \mu_m^{(\varepsilon)}: A_i^{(z)}\longrightarrow A_i^{(z)} + J_i -\widetilde{J}_i.
\end{equation}
One can then prove the following relations:
\begin{align}
    &\mu_0^{(-1)} \mu_m^{(+1)} \mu_0^{(-1)} = \mu_{-m}^{(+1)},\\
    &\mu_m^{(+1)}\mu_n^{(+1)} = \mu_{m+n}^{(+1)},\\
    &\mu_m^{(-1)}\mu_m^{(-1)} = \mu_0^{(+1)},
\end{align} 
where $m$ is implied to be a cyclic integer variable.
These relations define the structure of a dihedral group $\mathrm{D}_6$ with $\mu_0^{(+1)}$ serving as the identity element.
The $\mu_m^{(+1)}$ transformations in this case are equivalent to rotations and $\mu_0^{(-1)}$ is the reflection operation.
Since $\mu^{\eta}$ and $\mu_m^{(\varepsilon)}$ commute, the most general duality transformation is written as $\mu^{(\eta)}\mu_m^{(\varepsilon)}$ leading to group structure equivalent to $\mathrm{D}_{6h}$.
Thus, we obtain a very interesting situation where the group of self-dualities is non-Abelian and is isomorphic to the point group of the underlying lattice.

\subsection{\label{subsec:duality_anis}Intermediate coupling case}

When the SOC is sufficiently strong, the coupling matrix $\boldsymbol{\mathcal{A}}_{ij}(\boldsymbol{\delta})$ takes the form of Eq.~(\ref{eq:general_coupling_matrix}) and (\ref{eq:SIA_coupling_matrix}).
The symmetries of the Hamiltonian are reduced down to the paramagnetic group as seen in Fig.~\ref{fig:SOC_symmetry_diagram}.
This has no affect on the out-of-plane order parameters, and $\mu^{(\eta)}$ are still valid self-duality maps.
However, the two-dimensional $E_g^{(14)}$ and $E_u^{(11)}$ planar irreps each split into two one-dimensional irreps ($\{B_{1g}, B_{2g}\}$ and $\{A_{1u},A_{2u}\}$ respectively).
As a result, the dualities obtained from permutations of these irreps with other two-dimensional irreps are no longer exact.
Out of the twelve elements of $\mu_m^{(\varepsilon)}$, only four remain: $\mu_0^{(+1)}$, $\mu_1^{(-1)}$, $\mu_3^{(+1)}$, and $\mu_4^{(-1)}$. 
It is straight-forward to check that $\mu_0^{(+1)}$ and $\mu_4^{(-1)}$ leave the anisotropic parameters unchanged,

\begin{equation}
    \mu_0^{(+1)},\mu_4^{(-1)}: \begin{cases}K_\alpha\phantom{-}\longrightarrow K_\alpha,\\ A_1^{(xy)}\longrightarrow A_1^{(xy)},\\A_2^{(xy)}\longrightarrow A_2^{(xy)},\\A_4^{(xy)}\longrightarrow A_4^{(xy)},\end{cases}
\end{equation}
while $\mu_3^{(+1)}$ and $\mu_1^{(-1)}$ change the sign of the out-of-plane bond-dependent interactions,

\begin{equation}
    \mu_3^{(+1)},\mu_1^{(-1)}: \begin{cases}K_\alpha\phantom{-}\longrightarrow \phantom{-}K_\alpha,\\ A_1^{(xy)}\longrightarrow -A_1^{(xy)},\\A_2^{(xy)}\longrightarrow\phantom{-}A_2^{(xy)},\\ A_4^{(xy)}\longrightarrow-A_4^{(xy)}.\end{cases}
\end{equation}
In addition to these transformations, global $C_4$ rotations around the $z$-axis now lead to a valid self-duality map $\zeta$, since they interchange the one-dimensional planar irreps, thus flipping the sign of the anisotropic interactions:

\begin{equation}
    \zeta: \begin{cases}K_\alpha\phantom{-}\longrightarrow -K_\alpha,\\ A_1^{(xy)}\longrightarrow -A_1^{(xy)},\\A_2^{(xy)}\longrightarrow -A_2^{(xy)},\\ A_4^{(xy)}\longrightarrow -A_4^{(xy)}.\end{cases}
\end{equation}
The elements $\{\mu_0^{(+1)},\mu_3^{(+1)},\mu_1^{(-1)},\mu_4^{(-1)}\}$ and $\zeta$ commute (even though the corresponding spin transformation matrices generally do not), and all together they form the Burnside group $B(3,2)$ $\mathbb{Z}_2\otimes \mathbb{Z}_2\otimes \mathbb{Z}_2$, which is Abelian.
Combining it with the $\mu^{(\eta)}$, the group of the self-duality transformations in the intermediate SOC limit becomes the Burnside group $B(4,2)$.

\subsection{\label{subsec:duality_effects}Consequences of self-duality}

The consequences of self-duality are far-reaching.
In general, self-dual transformations can be thought of as the symmetries of the parameter space that do not change the energy of the system.
As a result, the dual points in the parameter space (self-dual images) will have a lot of the same physical properties.
These include the ground state energy and most thermodynamic properties, since the self-duality applies also to the partition function.
These results are also not limited to classical systems either: the self-dual images must have mostly the same quantum properties, within redefinition of quantization axes.
Therefore, given a description of a single phase, one can immediately describe all phases that relate to it via self-duality transformations. 
In the following, we discuss a few important properties of self-duality relevant to this work. 

From the properties of the duality relations it follows that in order to prove that a certain spin structure is stable for some choice of physical parameters, it is sufficient to show that one of its images is stable somewhere in the parameter space.
Given that even in the simple case of weak SOC limit the parameter space is four-dimensional, the self-duality maps can significantly reduce the time of computations since for every phase boundary $f(\boldsymbol{\mathcal{A}})$, an image $f\left(\mu\left(\boldsymbol{\mathcal{A}}\right)\right)$ must also correspond to a phase boundary.
At the phase boundaries, self-dualities become symmetry operations, and the Hamiltonian remains invariant after the transformation.
\begin{table}[!t]
    \centering
    \resizebox{0.5\textwidth}{!}{\begin{tabular}{c|r|r|r|r|r|r|r|r}
        $m$ &$\widetilde{J}_1$ & $\widetilde{D}_1$ & $\widetilde{A}^{(z)}_1$ & $\widetilde{J}_2$ & $\widetilde{D}_2$ & $\widetilde{A}^{(z)}_2$ & $\widetilde{J}_4$ & $\widetilde{A}^{(z)}_4$ \\
        \hline
        0 & $J_1$ & 0 & 0 & $J_2$ & 0 & 0 & $J_4$ & 0 \\
        1 & $\frac{1}{2}J_1$ & $\frac{\sqrt{3}}{2}J_1$ & $-\frac{1}{2}J_1$ & $-\frac{1}{2}J_2$ & $\frac{\sqrt{3}}{2}J_2$ & $-\frac{3}{2}J_2$ & $-J_4$ & $-2J_4$\\
        2 & $-\frac{1}{2}J_1$ & $\frac{\sqrt{3}}{2}J_1$ & $-\frac{3}{2}J_1$ & $-\frac{1}{2}J_2$ & $-\frac{\sqrt{3}}{2}J_2$ & $-\frac{3}{2}J_2$ & $J_4$ & 0\\
        3 & $-J_1$ & 0 & $-2J_1$ & $J_2$ & 0 & 0 & $-J_4$ & $-2J_2$ \\
        4 & $-\frac{1}{2}J_1$ & $-\frac{\sqrt{3}}{2}J_1$ & $-\frac{3}{2}J_1$ & $-\frac{1}{2}J_2$ & $\frac{\sqrt{3}}{2}J_2$ & $-\frac{3}{2}J_2$ & $J_4$ & 0\\
        5 & $\frac{1}{2}J_1$ & $-\frac{\sqrt{3}}{2}J_1$ & $-\frac{1}{2}J_1$ & $-\frac{1}{2}J_2$ & $-\frac{\sqrt{3}}{2}J_2$ & $-\frac{3}{2}J_2$ & $-J_4$ & $-2J_4$
    \end{tabular}}
    \caption{Points of accidental $\mathcal{G}_J$ symmetry obtained via self duality maps $\mu_m^{(\varepsilon)}$. Note that the value of $\varepsilon$ does not change the number of points.}
    \label{tab:accidental_degeneracy}
\end{table}

Self-dualities also allow one to quickly find the points of accidental degeneracy. 
Consider, for example, Eq.~(\ref{eq:duality_weak_SOC}) where the original DM parameters are set to zero $D_1=D_2=0$.
We get a set of new parameters, where $\widetilde{D}_1$ and $\widetilde{D}_2$ are generally non-zero.
However, since the self-duality preserves the transformation properties and therefore the symmetry group of the Hamiltonian, all self-dual images in this case must be completely isotropic in spin space (\textit{i.e.} have spin group $\mathcal{G}_J$).
Note that as stated before, in addition to the $\widetilde{J}_k$ and $\widetilde{D}_k$ one must also specify $\widetilde{A}^{(z)}_k = \widetilde{J}_k-J_k$ to ensure that the full rotational symmetry is exact.
With this, we get a set of points with accidental $\mathcal{G}_J$ degeneracy, listed in table~\ref{tab:accidental_degeneracy}.
At these points we expect the spin-wave dispersion to obtain additional zero-energy \textit{pseudo}-Goldstone modes. 

It is important to note, however, that the properties which depend directly on the spin structure (\textit{e.g.} average magnetization) are not always the same, since the self-duality maps generally do not conserve the order parameter.
An important consequence of this is that the self-dual images will have the same spin wave excitation spectra only when the corresponding spin transformations produce proper local axes (\textit{i.e.} when the local transformations are rotations).
This will be discussed in more detail in Sec.~\ref{sec:spin_waves}.

\section{\label{sec:phase_diagram}Classical Phase Diagrams for Decoupled and Weakly Coupled Cases}

\subsection{\label{subsec:pd_overview}Brief overview of the results}

\begin{figure}[!ht]
    \centering
    \includegraphics[width=0.45\textwidth]{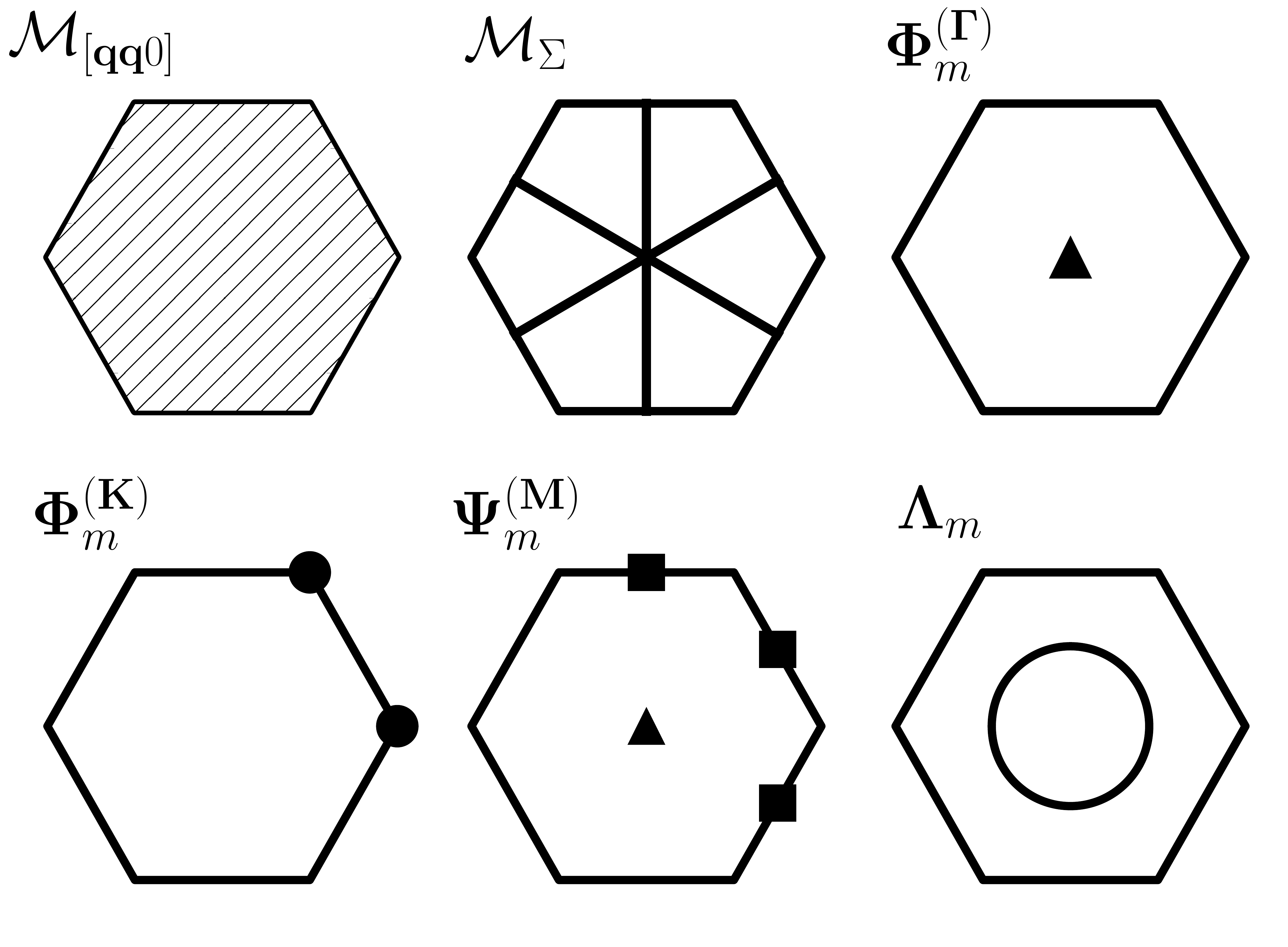}
    \caption{Magnetic ordering wavevectors corresponding to phases studied in this paper shown in the first Brillouin zone of an AB-stacked kagome lattice. $\mathcal{M}_{[\mathbf{q}\mathbf{q}0]}$ and $\mathcal{M}_\Sigma$ represent degenerate ground state manifolds. $\boldsymbol{\Phi}_m^{(\mathbf{Q})}$ and $\boldsymbol{\Psi}_m^{(\mathbf{M})}$ label the single- and multi-$\mathbf{Q}$ states, and $\boldsymbol{\Lambda}_m$ labels the incommensurate phases with Ising-type order, as described in the main text.}
    \label{fig:orders_BZ}
\end{figure}

In this section we present the classical magnetic phases for models with either negligible or weak SOC.
The effects of intermediate SOC are discussed in Sec.~\ref{sec:anisotropy}.
Before starting a detailed discussion of magnetic ground states, we summarize the main results of this section.
Based on our findings, we group the magnetic structures into three categories: single-$\mathbf{Q}$ configurations, multi-$\mathbf{Q}$ structures, and incommensurate phases with Ising-like ordering.
We label these as $\boldsymbol{\Phi}_m^{(\mathbf{Q})}$, $\boldsymbol{\Psi}_m^{(\mathbf{M})}$, and $\boldsymbol{\Lambda}_m$ respectively.
The integer label $m$ here refers to the different self-dual images produced by the $\mu_m^{(\varepsilon)}$ elements in the weak SOC limit, as discussed in the previous section. 
In addition to these states, in the decoupled limit, some phase boundaries correspond to structures with degenerate wavevectors in the Brillouin zone.
The magnetic ordering wavevectors of all observed states are shown in Fig.~\ref{fig:orders_BZ}.

\begin{figure*}[t]
    \centering
    \includegraphics[width=0.93\textwidth]{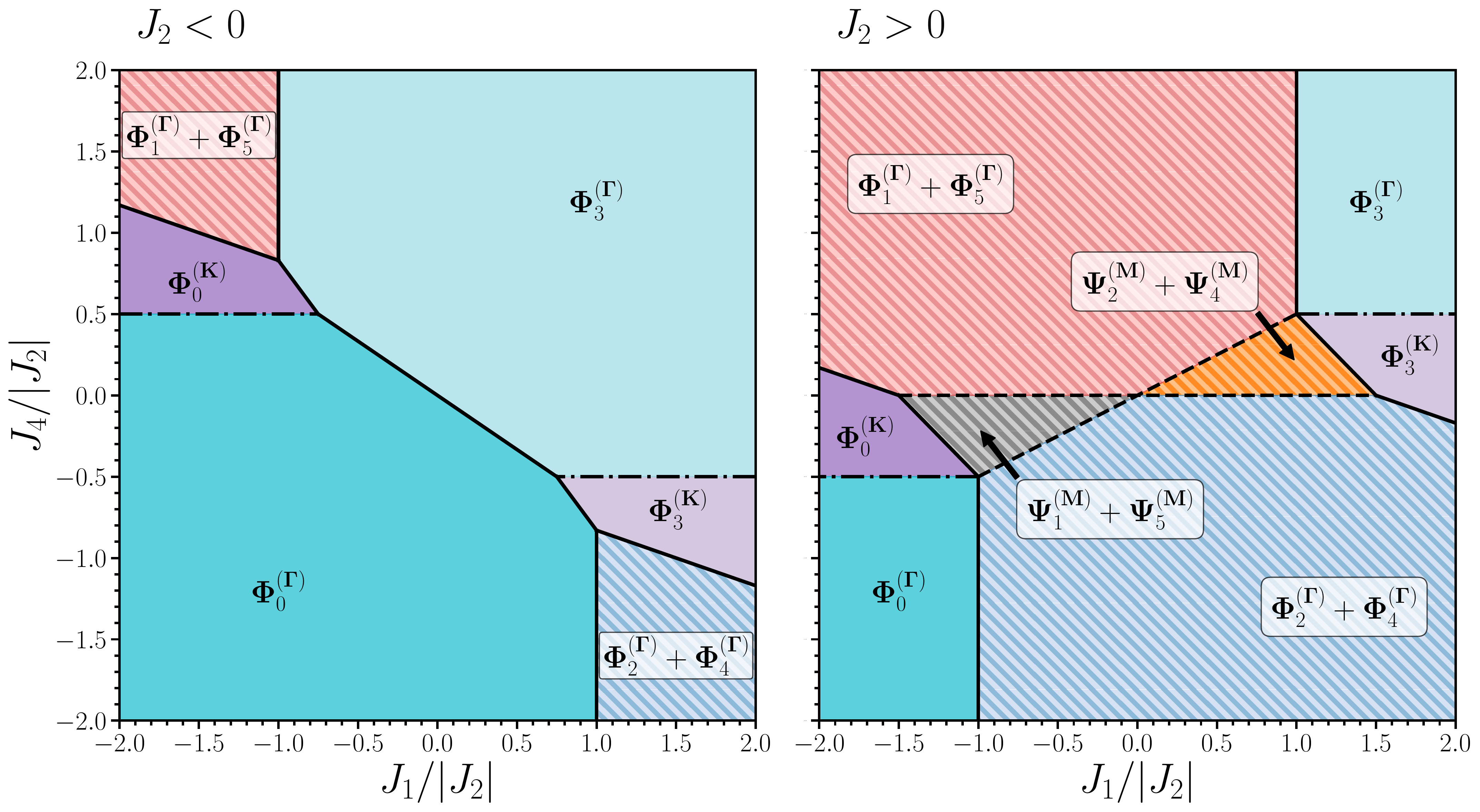}
    \caption{Ground state phase diagram for exchange-coupled spins on AB-stacked kagome lattice. Here, the dashed and dash-dotted boundaries correspond to the $\mathcal{M}_\Sigma$ and $\mathcal{M}_{[\mathbf{q}\mathbf{q}0]}$ manifold states. Since the exchange interactions do not differentiate between the different chirality values, the $m=1,5$ ($m=2,4$) states are written as linear combinations.}
    \label{fig:pd_j1_j4_h}
\end{figure*}

LT calculations were found to give correct spin configurations and energies for the single-$\mathbf{Q}$ phases.
For the remaining types of phases, the analytical results gave non-normalized spin configurations and, as a result, lower energies compared to the MC simulations.
However, in most cases, LT provided a decent estimation of the locations of the phase boundaries, allowing us to optimize our numerical boundary search.
Therefore, the exact phase boundaries between $\boldsymbol{\Phi}_m^{(\mathbf{Q})}$-type states were calculated using the LT method, and the locations of the remaining boundaries were determined via numerical MC simulations.

We show that Heisenberg exchange interactions are sufficient to stabilize both single- and multi-$\mathbf{Q}$ phases.
The inclusion of DM interactions lifts the degeneracy of the non-collinear phases by stabilizing configurations with particular chirality, which further enriches the magnetic phases. 
For most values of exchange constants, the $D_1 - D_2$ phase diagrams display multiple non-trivial ($\mathbf{Q}\neq 0$) structures, including unconventional states with Ising-like ordering.

\subsection{\label{subsec:pd_exchange}Decoupled case}

\begin{figure*}[!t]
    \centering
    \includegraphics[width=\textwidth]{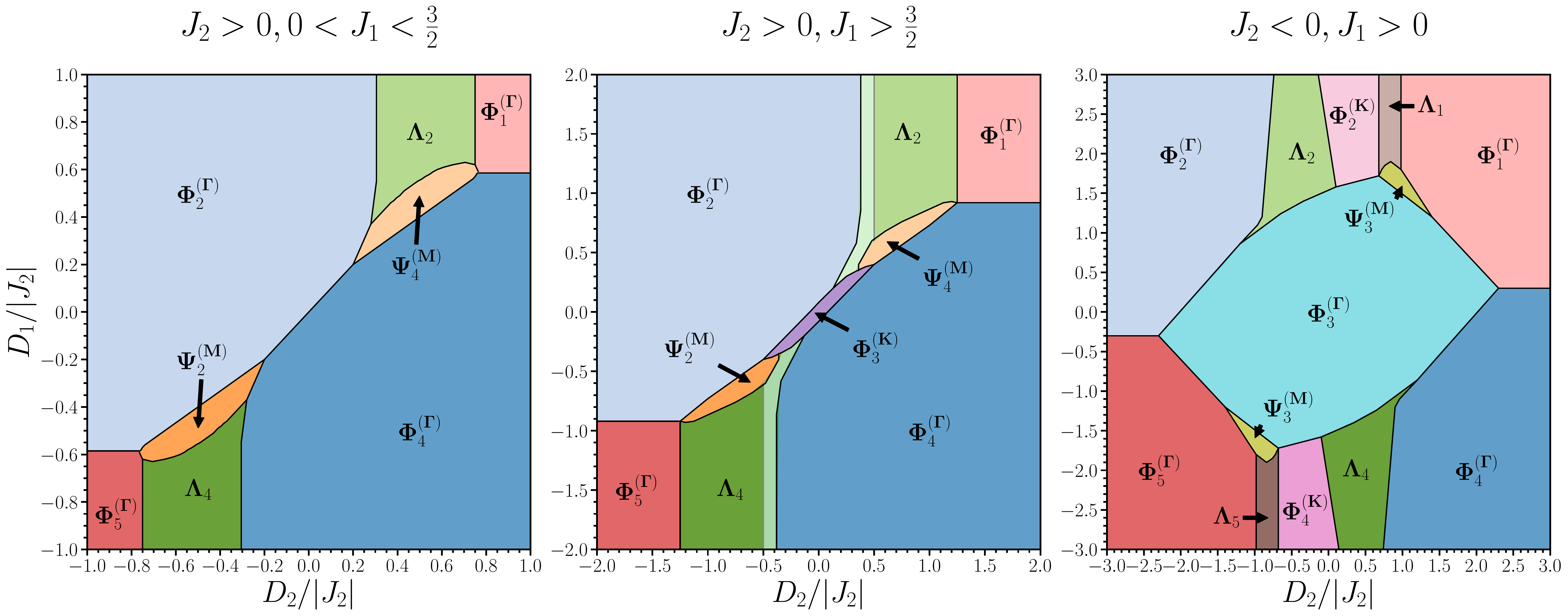}
    \caption{Ground state phase diagrams on AB-stacked kagome lattice for selected values of exchange and a range of DM values. In the middle diagram, the brightened strips between the $\boldsymbol{\Lambda}_{2(4)}$ and $\boldsymbol{\Phi}_{2(4)}^{(\boldsymbol{\Gamma})}$ phases are coexistence regions with many stable multi-\textbf{Q} configurations. }
    \label{fig:pds}
\end{figure*}

First, we present phase diagrams for the model in Eq.~(\ref{eq:magnetic_hamiltonian}) with $\mathcal{H} = \mathcal{H}_J$ in Fig.~\ref{fig:pd_j1_j4_h}.
Both diagrams ($J_2>0$ and $J_2<0$) display a clear ``inversion'' symmetry corresponding to flipping the sign of both $J_1$ and $J_4$, as a consequence of the self-duality map $\gamma^{(-1)}$, discussed in Sec.~\ref{subsec:duality_J}.
The results yield eight distinct phases, including single-, and multi-$\mathbf{Q}$ structures.
When $J_1$ and $J_4$ have the same sign and sufficiently large magnitudes, the ground state spin configurations become collinear, with spins in the A and B layers either parallel ($J_1<0$, $J_4<0$) or antiparallel ($J_1>0$, $J_4>0$) to each other.
On the other hand, opposing signs of the interactions introduce frustration, which for large magnitudes of the couplings lead to uniform 120-degree states.
Notably, the exchange interactions alone do not differentiate between states with different chirality, leaving the structures with $m=1,5$ ($m=2,4$) degenerate.
Interestingly enough, for sufficiently large magnitudes of $J_1$ there is a range of $J_4$ values that stabilizes commensurate cycloidal configurations regardless of the sign of $J_2$.
These structures are characterized by a period three modulation ($\mathbf{Q}=K$) with collinear spins in each unit cell (Fig.~\ref{fig:qK_phases}).
The stability of these phases extends indefinitely for large magnitudes of $J_1$. 
In the case of antiferromagnetic nn in-plane interactions, one can also stabilize multi-$\mathbf{Q}$ configurations for intermediate values of $J_1$ and $J_4$. 
Similar to the uniform 120-degree phases, the $\boldsymbol{\Psi}_m^{(\mathbf{M})}$ structures in Fig.~\ref{fig:pd_j1_j4_h} correspond to degenerate configurations with opposite chiralities. 

We also note that at certain phase boundaries structures with degenerate wavevectors can be stabilized.
In particular, the boundaries between $\boldsymbol{\Phi}_0^{(\boldsymbol{\Gamma})}$ ($\boldsymbol{\Phi}_3^{(\boldsymbol{\Gamma})}$) and $\boldsymbol{\Phi}_0^{(\mathbf{K})}$ ($\boldsymbol{\Phi}_3^{(\mathbf{K})}$) states stabilize structures where all wavevectors with $Q_z=0$ are degenerate, leading to a two-dimensional ground state manifold $\mathcal{M}_{[\mathbf{q}\mathbf{q}0]}$ (dash-dotted lines in Fig.~\ref{fig:pd_j1_j4_h}).
Also, the boundaries between $\boldsymbol{\Phi}_m^{(\boldsymbol{\Gamma})}$ and $\boldsymbol{\Psi}_m^{(\mathbf{M})}$ are degenerate along the $\Sigma$ lines in the Brillouin zone and are labeled as $\mathcal{M}_\Sigma$ (dashed lines in Fig.~\ref{fig:pd_j1_j4_h}).
In this work we will focus more on the ordered states and leave the description of the degenerate manifold states outside of the scope of this work.

\subsection{\label{subsec:pd_dmi}Weak coupling case}

Next, we consider the effects of DM interactions.
To reduce the number of independent parameters, we focus on the case where $J_4=0$, and present the diagrams that display the majority of the magnetic phases observed in this study.
The remaining phases can be obtained from the self-duality operations $\mu_m^{\varepsilon}$ presented in Sec.~\ref{subsec:duality_JD} [\textbf{Supplemental Material}]. 

The phase diagrams are shown in Fig.~\ref{fig:pds}.
We see that DM interactions break the degeneracy associated with the different chirality values of the non-collinear spin structures.
For large magnitudes of both $D_1$ and $D_2$, the ground state eventually becomes one of the $\mathbf{Q}=\Gamma$ 120-degree structures.
For the intermediate values of these constants, the competition between the exchange and DM interactions introduces a plethora of unconventional magnetic phases.
These include single-$\mathbf{Q}$ phases and multi-$\mathbf{Q}$ configurations already discussed, as well as incommensurate phases that manifest themselves as Ising-like stripes.
The latter are stable for a wide range of the DM parameters and extend far beyond the ranges shown in Fig.~\ref{fig:pds}.
These states were found to be hard to resolve in the simulations and required a careful choice of parameters, as well as many MC steps in order to obtain ordered stripe configurations.
Furthermore, for the $J_2>0$ $J_1/|J_2|>\frac{3}{2}$ diagram, we identified thin coexistence regions between the $\boldsymbol{\Lambda}_{2(4)}$ and $\boldsymbol{\Phi}_{2(4)}^{(\boldsymbol{\Gamma})}$ phases where various multi-$\mathbf{Q}$ configurations are stable.

\section{\label{sec:phases_structure}Structure of the magnetic phases}

\subsection{\label{subsec:structure_single_q}Single-$\mathbf{Q}$ phases}

\begin{figure}[!th]
    \centering
    \includegraphics[width=0.45\textwidth]{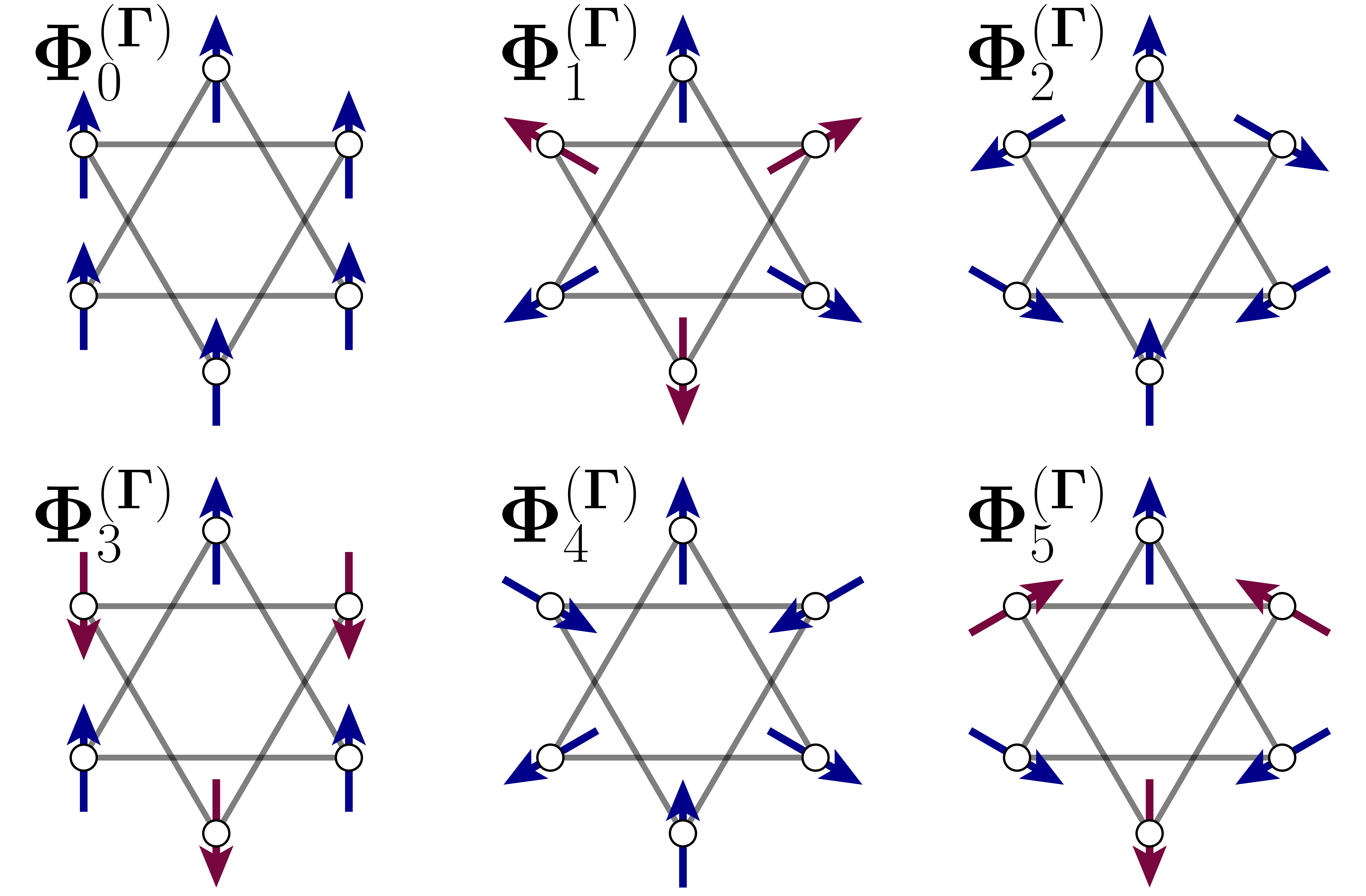}
    \caption{The $\boldsymbol{\Phi}^{(\boldsymbol{\Gamma})}_m$ spin configurations. Blue and red arrows are used to differentiate the structures that are symmetric and antisymmetric under the spatial inversion.}
    \label{fig:q0_phases}
\end{figure}

\begin{figure}[!t]
    \centering
    \includegraphics[width=0.5\textwidth]{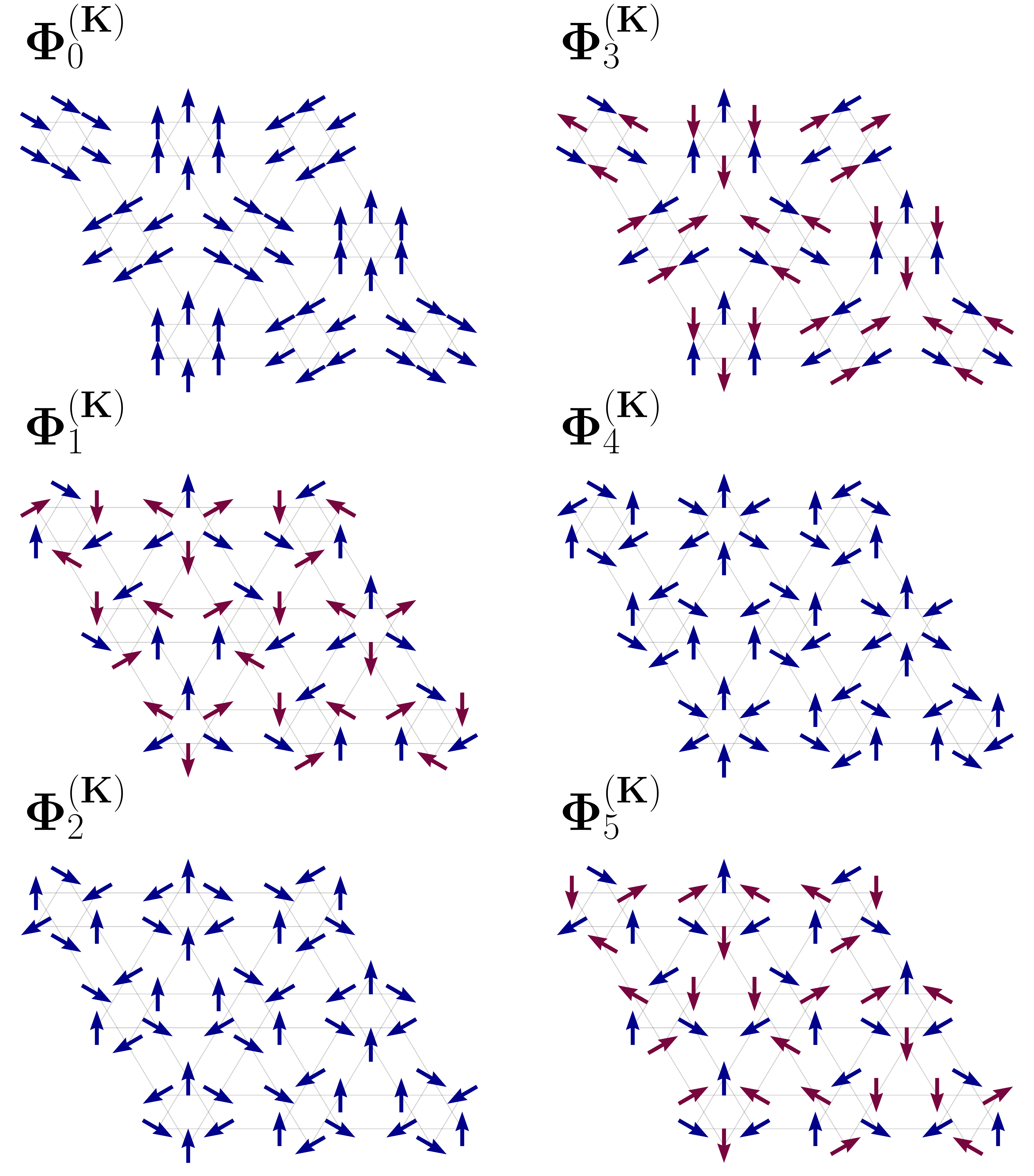}
    \caption{The $\boldsymbol{\Phi}^{(\mathbf{K})}_m$ spin configurations.}
    \label{fig:qK_phases}
\end{figure}

In this work, we encounter two types of single-$\mathbf{Q}$ phases: those with $\mathbf{Q}=\Gamma$ and $\mathbf{Q}=K$, which we label as $\boldsymbol{\Phi}^{(\boldsymbol{\Gamma})}_m$ and $\boldsymbol{\Phi}^{(\mathbf{K})}_m$ respectively.
The spin configurations of these phases are shown in figures~\ref{fig:q0_phases} and \ref{fig:qK_phases}.
Note that for exchange-only interactions, the spins generally also possess out-of-plane components, since the order parameters in this case belong to one of the four irreps in Sec.~\ref{subsec:symmetry_decoupled}.

When a weak SOC is turned on in the system, the in-plane and out-of-plane components are no longer equivalent and correspond to different irreps (Sec.~\ref{subsec:symmetry_weakly_coupled}).
Thus, in the decoupled limit, these structures can be parameterized as $\mathbf{S}_i(\mathbf{r}) = R(\phi_z,\phi)R^{(\mathbf{Q})}(\mathbf{r})M_i^{(m)} \mathbf{\hat{S}}$, where $\mathbf{\hat{S}}$ is an arbitrary in-plane unit vector, $M_i^{(m)}$ is defined in Eq.~(\ref{eq:matrix_dual_m}), and the remaining two rotation matrices are defined as
\begin{align}
    R(\phi_z,\phi) &= \begin{bmatrix} \cos(\phi) & -\sin(\phi) & 0\\\cos(\phi_z)\sin(\phi) & \cos(\phi_z)\cos(\phi) &-\sin(\phi_z)\\ 
    \sin(\phi_z)\sin(\phi) & \sin(\phi_z)\cos(\phi) & \cos(\phi_z)\end{bmatrix},
    \label{eq:single_Q_param_z}\\
    R^{(\mathbf{Q})}(\mathbf{r}) &= \begin{bmatrix} \cos(\mathbf{Q}\cdot\mathbf{r}) &-\sin(\mathbf{Q}\cdot\mathbf{r}) & 0\\ 
    \sin(\mathbf{Q}\cdot\mathbf{r}) & \cos(\mathbf{Q}\cdot\mathbf{r}) & 0\\
    0 & 0 & 1\end{bmatrix},
    \label{eq:single_Q_param_r}
\end{align} 
Matrix $R(\phi_z,\phi)$ is used to define the globally phase of the spin configurations.
In the weakly-coupled limit, the parameterization is changed to $\mathbf{S}_i(\mathbf{r}) = R(0,\phi)R^{(\mathbf{Q})}(\mathbf{r})M_i^{(m)} \mathbf{\hat{S}}$.
The corresponding classical energies of these spin configurations are presented in table~\ref{tab:single_Q_energies}.
\begin{table}[h]
    \centering
    \begin{tabular}{c|c|c}
        $m$ & $\mathcal{H}^{0}\left(\boldsymbol{\Phi}_m^{(\boldsymbol{\Gamma})}\right)$ & $\mathcal{H}^{0}\left(\boldsymbol{\Phi}_m^{(\mathbf{K})}\right)$ \\
        \hline
        0 & $2J_1+2J_2+3J_4$ & $2J_1+\frac{J_2}{2}$ \\
        1 & $J_1-J_2-3J_4-\sqrt{3}(D_1+D_2)$ & $J_1-\frac{J_2}{4}-\sqrt{3}(D_1+\frac{D_2}{4})$ \\
        2 & $-J_1-J_2+3J_4-\sqrt{3}(D_1-D_2)$ & $-J_1-\frac{J_2}{4}-\sqrt{3}(D_1-\frac{D_2}{4})$ \\
        3 & $-2J_1+2J_2-3J_4$ & $-2J_1+\frac{J_2}{2}$ \\
        4 & $-J_1-J_2+3J_4+\sqrt{3}(D_1-D_2)$ & $-J_1-\frac{J_2}{4}+\sqrt{3}(D_1-\frac{D_2}{4})$ \\
        5 & $J_1-J_2-3J_4+\sqrt{3}(D_1+D_2)$ & $J_1-\frac{J_2}{4}+\sqrt{3}(D_1+\frac{D_2}{4})$
    \end{tabular}
    \caption{Classical energies of the single-$\mathbf{Q}$ configurations.}
    \label{tab:single_Q_energies}
\end{table}
We note that while formally the $z$-components of spins are decoupled from the planar ones in the weak SOC limit, the energy of the $\boldsymbol{\Phi}_0^{(\mathbf{Q})}$ and $\boldsymbol{\Phi}_3^{(\mathbf{Q})}$ structures is independent of the DM parameters leading to a full rotational degeneracy at the classical level.
This degeneracy originates from the collinear sublattice structure in these phases. 
As will be discussed in Sec.~\ref{sec:spin_waves}, thermal fluctuations break this classical degeneracy and orient the structures in the out-of-plane direction. 

\subsection{\label{subsec:structure_multi_q}Multi-$\mathbf{Q}$ phases}

\begin{figure}[!ht]
    \centering
    \includegraphics[width=0.41\textwidth]{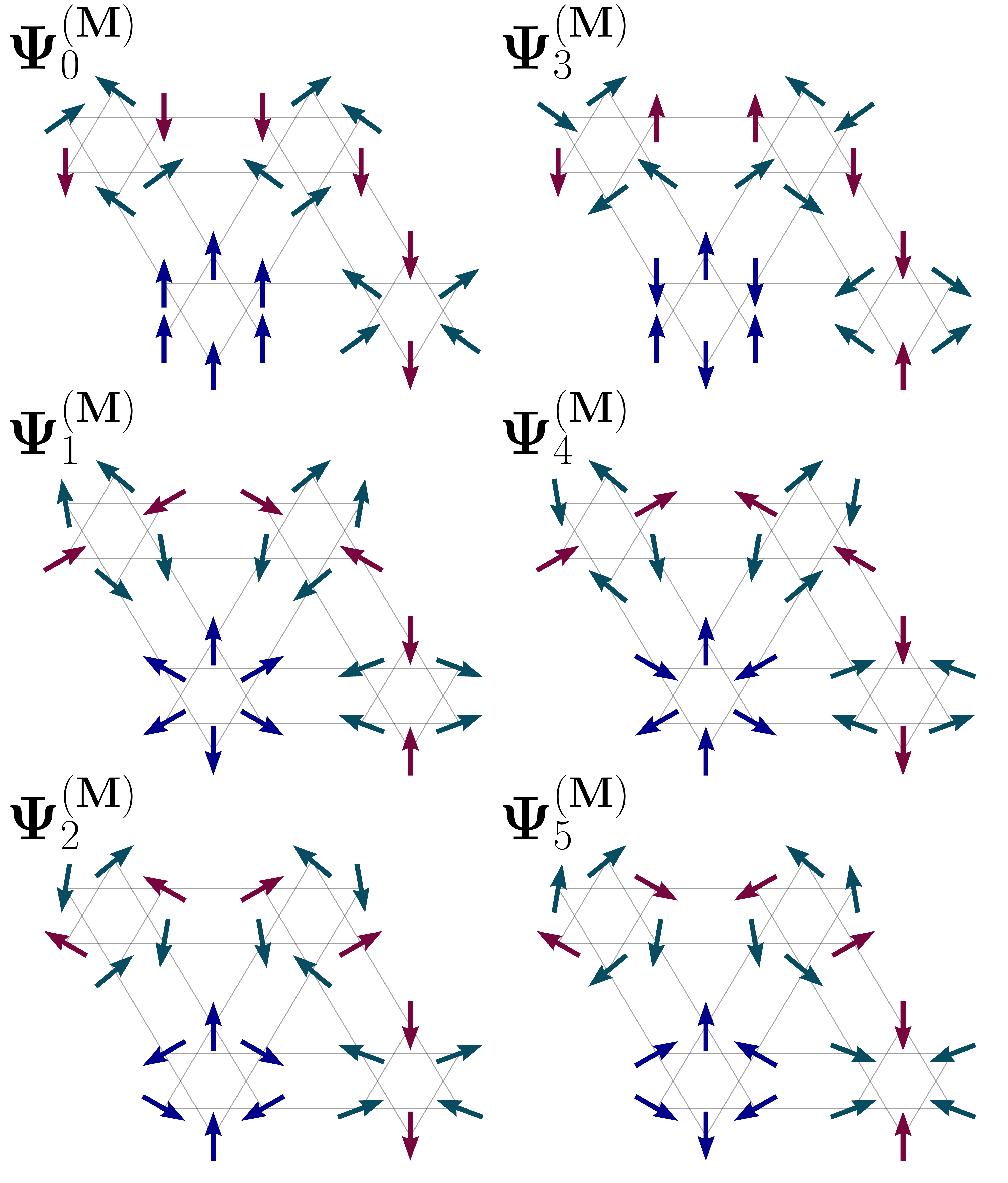}
    \caption{Magnetic unit cells of the $\boldsymbol{\Psi}_m^{(\mathbf{M})}$ configurations. Colors indicate spins that belong to the same orbit (see main text).}
    \label{fig:4q_phases}
\end{figure}

The most prominent type of multi-$\mathbf{Q}$ states in the phase diagrams in Fig.~\ref{fig:pd_j1_j4_h} and~\ref{fig:pds} is $\boldsymbol{\Psi}_m^{(\mathbf{M})}$.
The structure factor of these structures displays one peak at the $\Gamma$ point, as well as three peaks at the $M$ points in the first Brillouin zone (forming the star of $M$), as shown in Fig.~\ref{fig:orders_BZ}.
The spin configurations are shown in Fig.~\ref{fig:4q_phases}.
The parameterization of these structures is not trivial.
To make some further progress, we must consider the symmetry of these configurations in order to reduce the number of independent variables.
The analysis below is performed assuming the weak coupling case and then generalized to also describe the decoupled case. 

We first note that the peaks of the structure factor in Fig.~\ref{fig:orders_BZ} are completely symmetric with respect to the $\mathrm{D}_{6h}$ point group operations.
Therefore, even though these spin structures break planar translations by a single unit cell, the point group lattice transformations around the (0,0) site are preserved, provided that we combine them with the necessary spin rotations.
Thus, we can still express the $\boldsymbol{\Psi}_m^{(\mathbf{M})}$ configurations in terms of the two-dimensional $E_g^{(1m)}$ and $E_u^{(1m)}$ irreps of $\mathcal{G}_D$ in table~\ref{tab:GD_character_table}.
The general representation of the spin structures in Fig.~\ref{fig:4q_phases} decomposes as

\begin{equation}
    \boldsymbol{\Psi}^{(\mathbf{M})} = \bigoplus_m 4E_{a_m}^{(1m)},
\end{equation}
where $a_m = g$ if $m$ is even and $a_m = u$ otherwise.
Because all of the symmetry operations transform spins globally (\textit{i.e.} no site-dependent transformations), it is impossible to have spin configuration described by irreps with more than one value of the winding $m$, without inevitably breaking the $\mathrm{D}_{6h}$ lattice symmetry.
Therefore, a given spin configuration must decompose into four irreps with the same winding number $m$, $\boldsymbol{\Psi}_m^{(\mathbf{M})} = 4E_1^{(1m)}$.
As a result, the parameterization of $\boldsymbol{\Psi}_m^{(\mathbf{M})}$ can be written in terms of four planar vectors, which describe the spins belonging to the same \textit{orbit}.
Here, we define an orbit as a set of all spin components related to each other via lattice point group transformations (Fig.~\ref{fig:4q_phases}).
The four orbits are given below.
\begin{alignat}{2}
    \mathbf{O}^{(1)}_i &: \big\{&&\mathbf{S}_1(0,0),\mathbf{S}_2(0,0),\mathbf{S}_3(0,0),\notag\\
    & &&\mathbf{S}_4(0,0),\mathbf{S}_5(0,0),\mathbf{S}_6(0,0)\big\},\\[5pt]
    \mathbf{O}^{(2)}_i &: \big\{&&\mathbf{S}_1(1,0),\mathbf{S}_2(1,1),\mathbf{S}_3(0,1),\notag\\
    & &&\mathbf{S}_4(1,0),\mathbf{S}_5(1,1),\mathbf{S}_6(0,1)\big\},
\end{alignat}
\begin{alignat}{2}
    \mathbf{O}^{(3)}_i,\mathbf{O}^{(4)}_i &: \big\{&&\{\mathbf{S}_1(1,1), \mathbf{S}_2(0,1), \mathbf{S}_3(1,0), \notag\\
    & &&\mathbf{S}_4(1,1), \mathbf{S}_5(0,1), \mathbf{S}_6(1,0)\}, \notag\\
    & &&\{\mathbf{S}_1(0,1), \mathbf{S}_2(1,0), \mathbf{S}_3(1,1), \notag\\
    & &&\mathbf{S}_4(0,1), \mathbf{S}_5(1,0),  \mathbf{S}_6(0,1)\}\big\}.
\end{alignat}
The positions of the spins inside of the magnetic unit cell are given by two integers $(k,l)$, which correspond to displacement vector $\mathbf{r} = k\mathbf{a}_1+ l\mathbf{a}_2$. 
The last two orbits include different components of spins on the same sites.
Thus, the $\boldsymbol{\Psi}_m^{(\mathbf{M})}$ order parameters can generally be written in terms of the four vectors $\mathbf{O}_i^{(k)}$.
However, due to the constraint on the norm of the spin vectors, the orbits $\mathbf{O}^{(3)}_i$ and $\mathbf{O}^{(4)}_i$ are no longer independent and will share the parameterization parameters.
We can therefore write 

\begin{equation}
    \boldsymbol{\Psi}_m^{(\mathbf{M})} = \min_{\phi_k}\sum_{k=1}^4 R(0,\phi_k)M_i^{(m)}\mathbf{O}_i^{(k)},
\end{equation} where $M_i^{(m)}$ is given in Eq.~(\ref{eq:matrix_dual_m}), $R(\phi_z,\phi)$ is defined in Eq.~(\ref{eq:single_Q_param_z}), and $\phi_k$ are the angles that parameterize the spin structure.
The exact spin configurations are then obtained by minimizing the energy with respect to the $\phi_k$. 
To simplify the minimization, we can remove the in-plane rotations that do not change the energy by fixing the value of $\phi_1$, and also use the equivalence of $\mathbf{O}_i^{(3)}$ and $\mathbf{O}_i^{(4)}$ to deduce that $\phi_3=-\phi_4$.
This leaves us with two independent variables which can be determined using numerical minimization.

Finally, we note that in the decoupled limit the above analysis still holds, although the spin configurations may also possess out-of-plane components, which can be parameterized by introducing a non-zero angle $\phi_z$.

\subsection{\label{subsec:structure_ising} Ising-like phases}
\begin{figure*}[!ht]
    \centering
    \includegraphics[width=\textwidth]{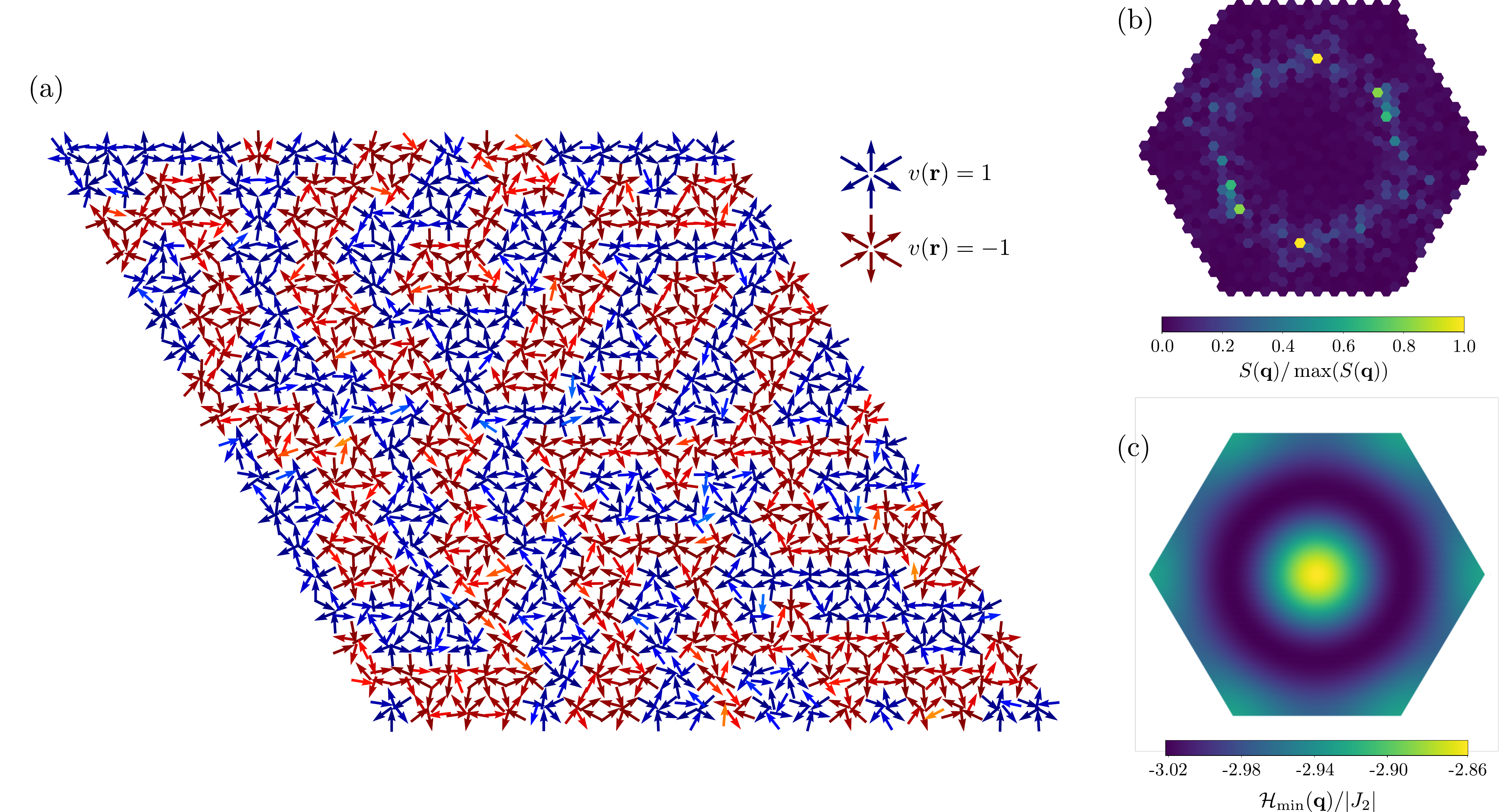}
    \caption{(a) A snapshot of the low-temperature $\boldsymbol{\Lambda}_4$ spin configuration on a lattice with $18^3\times 6$ spins. This state was obtained using $J_1 = 1$, $J_2 = 1$, $D_1 = -1$, and $D_2 = -0.5$. The order parameters that define the Ising variables are shown in the top right corner. (b) Structure factor calculated for a system with $30\times 30\times 2\times 6$ spins. The values were normalized with respect to the maxima. The parameters used in the simulation are the same as in (a). (c) Fourier transform of the effective Ising Hamiltonian in Eq.~(\ref{eq:corrected_energy}) obtained using the LT approximation.}
    \label{fig:ising_phases}
\end{figure*}

Lastly, we consider the $\boldsymbol{\Lambda}_m$ phases.
The complete description of these structures is complicated by the fact that there appears to be a large number of local minima that are very close in energy to the ground state configuration.
Furthermore, the spin configurations depend strongly on the size of the system, indicating an incommensurate nature of the magnetic order.
As a result, in order to resolve a single structure, it is necessary to perform long ($\sim 10^5$ MC steps) simulations on large ($> 3\cdot 10^4$ spins) systems.
The resulting $\boldsymbol{\Lambda}_m$ spin configurations often manifest in non-trivial patterns, where in each unit cell, the order parameter is approximately equal to $\pm\boldsymbol{\Phi}_m^{(\boldsymbol{\Gamma})}$.
The sign of this effective order parameter alternates rapidly throughout the system, giving rise to discrete domains.
For this reason, we refer to these states as Ising-like.

In the following, we will focus our attention on the description of this Ising behavior by considering the $\boldsymbol{\Lambda}_4$ phase, although the discussion equally applies to all $\boldsymbol{\Lambda}_m$ states by the virtue of the self-duality transformations (Sec.~\ref{sec:self_duality}).
Since these phases do not appear in the decoupled limit, the discussion below assumes weak SOC.
Fig.~\ref{fig:ising_phases}~(a) shows a typical low-temperature $\boldsymbol{\Lambda}_4$ spin configuration.
It is important to note that each kagome layer is identical, corresponding to $Q_z=0$, but within the plane, the pattern can often appear seemingly random.
The structure factor (Fig.~\ref{fig:ising_phases}~(b)) displays a characteristic ring in the Brillouin zone with a relatively well-defined radius.
We note in this ring there are typically a couple of peaks with larger intensity, corresponding to the dominant orientations of the stripy domains.
In our simulations, the spin deviations from the $\boldsymbol{\Phi}_4^{(\boldsymbol{\Gamma})}$ state in each unit cell were found to be restricted to the plane of the kagome (the out-of-plane deviations are typically four orders of magnitude smaller).
Therefore, for simplicity, we express the spins as two-dimensional vectors 

\begin{equation}
    \mathbf{S}_i(\mathbf{r}) = \begin{bmatrix}\cos\left( \theta_i^{(4)} + \delta\tilde{\theta}_i(\mathbf{r}) \right)\\\sin\left( \theta_i^{(4)} + \delta\tilde{\theta}_i(\mathbf{r}) \right)\end{bmatrix},
\end{equation} where $\theta_i^{(m)}$ are the angles in Eq.~(\ref{eq:duality_angles}) that correspond to the $\boldsymbol{\Phi}_4^{(\boldsymbol{\Gamma})}$ order parameter, and $\delta\tilde{\theta}_i(\mathbf{r})$ are the deviations from the order parameter.
In order to account for the alternating sign of the $\boldsymbol{\Phi}_4^{(\boldsymbol{\Gamma})}$ order parameter, we further rewrite these deviations as 

\begin{equation}
    \delta\tilde{\theta}_i(\mathbf{r}) = -\frac{v(\mathbf{r})-1}{2}\pi + \delta\theta_i(\mathbf{r}),
\end{equation} where $v(\mathbf{r})$ are Ising variables that take values $\pm 1$, and $0\leq\delta\theta_i(\mathbf{r})<\pi$ are the residual angle deviations.
Substituting this parameterization into the Hamiltonian with exchange and DM interactions is equivalent to performing a $\mu_4^{(+1)}$ gauge transformation, from which we obtain

\begin{align}
     \mathcal{H} = \sum_{\mathbf{r}\mathbf{r}'}\Bigg[\sum_{\langle ij\rangle} &\widetilde{J}_{ij}(\mathbf{r}-\mathbf{r}')\cos\Big(\delta\theta_{ij}(\mathbf{r},\mathbf{r}')\Big) \label{eq:JD_ising_local_coord}\\
     + &\widetilde{D}_{ij}(\mathbf{r}-\mathbf{r}')\sin\Big(\delta\theta_{ij}(\mathbf{r},\mathbf{r}')\Big)\Bigg]v(\mathbf{r})v(\mathbf{r}'),\notag
\end{align} where $\widetilde{J}_{ij}(\boldsymbol{\delta})$ and $\widetilde{D}_{ij}(\boldsymbol{\delta})$ are given in Eq.~(\ref{eq:duality_weak_SOC}) ($m=4$ and $\varepsilon=+1$), and $\delta\theta_{ij}(\mathbf{r},\mathbf{r}')=\delta\theta_i(\mathbf{r})-\delta\theta_j(\mathbf{r}')$.
Eq.~(\ref{eq:JD_ising_local_coord}) can be recast into an Ising model on a triangular lattice with non-local nn interactions:

\begin{equation}
    \mathcal{H} = \sum_{\mathbf{r}\mathbf{r}'}\mathcal{J}(\mathbf{r},\mathbf{r}')v(\mathbf{r})v(\mathbf{r}') + \mathcal{B}(\mathbf{r},\mathbf{r}').
\end{equation} Here, both the Ising exchange coupling $\mathcal{J}(\mathbf{r},\mathbf{r}')$, and constant $\mathcal{B}(\mathbf{r},\mathbf{r}')$ depend on $\delta\theta_i(\mathbf{r})$.

It is worth discussing the values of $\widetilde{J}_{ij}(\boldsymbol{\delta})$ and $\widetilde{D}_{ij}(\boldsymbol{\delta})$ for which the $\boldsymbol{\Lambda}_4$ is stable.
In fig~\ref{fig:ising_phases}, the parameters are $J_1=J_2 = 1$, $D_1=-1$, and $D_2=-0.5$, which corresponds to $\widetilde{J}_1 \approx -1.37$, $\widetilde{J}_2 \approx -0.07$, $\widetilde{D}_1 \approx -0.37$, and $\widetilde{D}_2 \approx 1.12$.
Therefore, in the local coordinate frame, the inter-plane interactions are dominated by the ferromagnetic exchange, whereas the in-plane interactions almost exclusively come from the DM term.
Notably, $\boldsymbol{\Lambda}_4$ phase in this case (Fig.~\ref{fig:pds}) appears to be stable for arbitrarily large values of $D_1$, which leads to large negative values of $\widetilde{J}_1$.

Using this information, we can expand Eq.~(\ref{eq:JD_ising_local_coord}) to second order in $\delta\theta_i(\mathbf{r})$, while assuming that $\widetilde{J}_2$ is of the same order of magnitude as the deviations.
We also impose ferromagnetic order along the $z$-direction by setting $\delta\theta_i(\mathbf{r}\pm \mathbf{a}_3) = \delta\theta_i(\mathbf{r})$ and $v(\mathbf{r}\pm \mathbf{a}_3) = v(\mathbf{r})$. Under these assumptions, Eq.~(\ref{eq:JD_ising_local_coord}) simplifies to the following quadratic equation:

\begin{equation}
    \mathcal{H} \approx \mathcal{H}^{(0)} + \sum_\mathbf{r}\sum_i g_i(\mathbf{r})\delta\theta_i(\mathbf{r}) + \frac{1}{2}\sum_\mathbf{r}\sum_{ij}H_{ij}\delta\theta_i(\mathbf{r})\delta\theta_j(\mathbf{r}),
    \label{eq:quadratic_form}
\end{equation} where $\mathcal{H}^{(0)}$ is independent of $\delta\theta_i(\mathbf{r})$
\begin{widetext}
\begin{align}
    \mathcal{H}^{(0)} = \sum_{\mathbf{r}} 6\widetilde{J}_1\Big[&1 + v(\mathbf{r})v(\mathbf{r}+\mathbf{a}_3) \Big] \notag\\+ 2\widetilde{J}_2\Big[&3 + v(\mathbf{r})v(\mathbf{r}+\mathbf{a}_1) +  v(\mathbf{r})v(\mathbf{r}+\mathbf{a}_2) + v(\mathbf{r})v(\mathbf{r}+\mathbf{a}_1+\mathbf{a}_2)\Big],
\end{align}
and $g_i(\mathbf{r})$ and $H_{ij}$ are the gradient vector and Hessian matrix respectively: 
\begin{equation}
    g_i(\mathbf{r}) = -\widetilde{D}_2v(\mathbf{r})\begin{bmatrix} v(\mathbf{r}-\mathbf{a}_2) - v(\mathbf{r}-\mathbf{a}_1-\mathbf{a}_2)\\ 
    v(\mathbf{r}-\mathbf{a}_1) - v(\mathbf{r}+\mathbf{a}_2)\\ 
    v(\mathbf{r}+\mathbf{a}_1+\mathbf{a}_2) - v(\mathbf{r}+\mathbf{a}_1)\\ 
    v(\mathbf{r}+\mathbf{a}_2) - v(\mathbf{r}+\mathbf{a}_1+\mathbf{a}_2)\\ 
    v(\mathbf{r}+\mathbf{a}_1) - v(\mathbf{r}-\mathbf{a}_2)\\
    v(\mathbf{r}-\mathbf{a}_1-\mathbf{a}_2) - v(\mathbf{r}-\mathbf{a}_1) \end{bmatrix},
    H_{ij} = 2\widetilde{J}_1\begin{bmatrix}-2 & 0 & 0 & 0 & 1 & 1\\
                                             0 &-2 & 0 & 1 & 0 & 1\\
                                             0 & 0 &-2 & 1 & 1 & 0\\
                                             0 & 1 & 1 &-2 & 0 & 0\\
                                             1 & 0 & 1 & 0 &-2 & 0\\
                                             1 & 1 & 0 & 0 & 0 &-2\end{bmatrix}.
    \label{eq:grad_hessian}
\end{equation} 
\end{widetext}
The Hessian matrix in Eq.~(\ref{eq:grad_hessian}) is positive semi-definite, with a single zero in the eigenvalue spectrum corresponding to global rotations.
After this mode is integrated out by fixing the planar orientations of the spins, a straight-forward minimization of the quadratic equation gives

\begin{equation}
    \delta\phi_i^\text{min}(\mathbf{r}) = -\sum_j H^{-1}_{ij}g_j(\mathbf{r}).
    \label{eq:solution_normal_quadratic_form}
\end{equation} 
Thus, the minimized energy can be written as

\begin{equation}
    \mathcal{H}_\text{min} = \mathcal{H}^{(0)} +dE,
    \label{eq:corrected_energy}
\end{equation}
where the energy contribution of the spin deviations is given by

\begin{widetext}
\begin{alignat}{2}
    dE &= -\frac{1}{2}\sum_\mathbf{r}\sum_{ij} H&&^{-1}_{ij}\mu_i(\mathbf{r})\mu_j(\mathbf{r}) \notag\\
    &= \phantom{-}\mathcal{K}\sum_\mathbf{r}v(\mathbf{r})\Bigg[ &&\phantom{-}2v(\mathbf{r}+\mathbf{a}_1) + 2v(\mathbf{r}+\mathbf{a}_2) + 2v(\mathbf{r}+\mathbf{a}_1+\mathbf{a}_2) \notag\\
    & &&+ v(\mathbf{r}+2\mathbf{a}_1) + v(\mathbf{r}+2\mathbf{a}_2) + v(\mathbf{r}+2\mathbf{a}_1+2\mathbf{a}_2)\notag\\
    & &&+ 2v(\mathbf{r}+\mathbf{a}_1-\mathbf{a}_2) + 2v(\mathbf{r}+2\mathbf{a}_1+\mathbf{a}_2) + 2v(\mathbf{r}+\mathbf{a}_1+2\mathbf{a}_2) -15 \Bigg],
    \label{eq:energy_contribution}
\end{alignat}
\end{widetext}
and the constant $\mathcal{K}$ is defined as
\begin{equation}
    \mathcal{K} = -\frac{\widetilde{D}_2^2}{12\widetilde{J}_1},
\end{equation}
Note that the positions of the Ising variables in Eq.~(\ref{eq:energy_contribution}) have been shifted to better indicate the nature of the effective interactions.  
We see that the spin deviations renormalize nearest-neighbor interactions between Ising variables and also induce second- and third-neighbor in-plane antiferromagnetic interactions (since $\widetilde{J}_1$ is negative).
These values of the coupling constants stabilize incommensurate phases in the triangular Ising antiferromagnets~\cite{Plumer_triangular_nnn_ising_1993_prb}.

Qualitatively \footnote{We note that a more systematic study of these phases is currently underway.}, we can understand the nature of these unusual phases in the following way.
The $\widetilde{J}_1$ interactions establish ferromagnetic ordering along the $z$-axis, ensuring $Q_z=0$.
$\widetilde{D}_2$ interactions couple small spin deviations to stabilize incommensurate in-plane wavevectors with a fixed magnitude $Q_I$.
LT analysis of equation~(\ref{eq:corrected_energy}) gives a dispersion with a degenerate ring with radius $Q_c$, in agreement with the MC results (Fig.~\ref{fig:ising_phases}~(c)). 
However, due to the Ising nature of the ordering, we expect there to be a larger degeneracy in the ground state than what is predicted by the LT method.
Although it is not evident from the analysis above, the phase diagrams in Fig.~\ref{fig:pds} indicate that the larger values of $\widetilde{J}_2$ serve to tune $Q_c$, until it either becomes zero ($\mathbf{Q}=\Gamma$) or reaches the zone boundary ($\mathbf{Q}=K,M$).

\section{\label{sec:spin_waves} Spin waves}

\subsection{\label{subsec:spin_waves_dynamic_matrix}Dynamical matrix}

In this section, we consider the elementary spin excitations in certain magnetic phases presented in Sec.~\ref{sec:phase_diagram}.
To do this, we solve the linearized torque equation~(\ref{eq:LL_equation}), as described in Sec.~\ref{subsec:methods_dynamics}.
Once the appropriate local coordinates are selected, the equations of motion can be generally written in the matrix form as
\begin{widetext}
\begin{equation}
    \frac{\text{d}}{\text{d}t}\begin{bmatrix}
    \widetilde{S}_{ix}(\mathbf{r},t)\\\widetilde{S}_{iy}(\mathbf{r},t)
    \end{bmatrix} = \sum_{\mathbf{r}'}\sum_j \begin{bmatrix}
    \widetilde{\mathcal{A}}_{yxij}(\mathbf{r}-\mathbf{r}')\widetilde{S}_{jx}(\mathbf{r}',t) + \widetilde{\mathcal{A}}_{yyij}(\mathbf{r}-\mathbf{r}')\widetilde{S}_{jy}(\mathbf{r}',t) - \widetilde{\mathcal{A}}_{zzij}(\mathbf{r}-\mathbf{r}')\widetilde{S}_{iy}(\mathbf{r},t)\\ \widetilde{\mathcal{A}}_{zzij}(\mathbf{r}-\mathbf{r}')S_{ix}(\mathbf{r}) - \widetilde{\mathcal{A}}_{xyij}(\mathbf{r}-\mathbf{r}')\widetilde{S}_{jy}(\mathbf{r}',t) - \widetilde{\mathcal{A}}_{xxij}(\mathbf{r}-\mathbf{r}')\widetilde{S}_{jx}(\mathbf{r}',t)
    \end{bmatrix}
    \label{eq:linearized_sw_time_space}
\end{equation}
\end{widetext}
where $\widetilde{\mathcal{A}}_{\alpha\beta ij}(\mathbf{r}-\mathbf{r}')$ are the elements of the coupling matrix in the local coordinates, $\mathbf{r}$ and $\mathbf{r}'$ determine the positions of the magnetic unit cells, and $i$ and $j$ generally label the non-equivalent magnetic sublattices.
Eq.~(\ref{eq:linearized_sw_time_space}) is solved by first defining the spatial Fourier transforms of the spin components as in Eq.~(\ref{eq:spin_FT}), and then Fourier transforming in the time domain ($\mathbf{S}_i(\mathbf{q},t) = \sum_{\omega} S_i(\mathbf{q},\omega)e^{-i\omega t}$).
The result is written as an eigenvalue equation:

\begin{equation}
    -i\omega(\mathbf{q})\begin{bmatrix}
    \mathbf{S}_{x}\\ \mathbf{S}_{y}
    \end{bmatrix} = \begin{bmatrix} \mathbf{u}^{(xx)}(\mathbf{q}) & \mathbf{u}^{(xy)}(\mathbf{q})\\ \mathbf{u}^{(yx)}(\mathbf{q}) & \mathbf{u}^{(yy)}(\mathbf{q}) \end{bmatrix}\begin{bmatrix}
    \mathbf{S}_{x}\\ \mathbf{S}_{y}
    \end{bmatrix}
    \label{eq:dynamical_matrix}
\end{equation}
where

\begin{align}
    u^{(xx)}_{ij}(\mathbf{q}) &= \widetilde{\mathcal{A}}_{yxij}(\mathbf{q})\\
    u^{(xy)}_{ij}(\mathbf{q}) &= \widetilde{\mathcal{A}}_{yyij}(\mathbf{q}) - \sum_k\widetilde{\mathcal{A}}_{yyjk}(\mathbf{q})\Delta_{ij}\\
    u^{(yx)}_{ij}(\mathbf{q}) &= \sum_k\widetilde{\mathcal{A}}_{zzjk}(\mathbf{q})\Delta_{ij} - \widetilde{\mathcal{A}}_{xxij}(\mathbf{q})\\
    u^{(yy)}_{ij}(\mathbf{q}) &= -\widetilde{\mathcal{A}}_{xyij}(\mathbf{q})
\end{align}
Here, $\Delta_{ij}$ is a Kronecker delta function, and $u^{(ab)}_{ij}(\mathbf{q})$ are the elements of the dynamical spin wave matrix.
For planar spin configurations, it is possible to choose local coordinates such that $u^{(xx)}_{ij}(\mathbf{q}) = u^{(yy)}_{ij}(\mathbf{q}) = 0$.
The frequencies $\omega(\mathbf{q})$ must generally be calculated numerically.
However, for some high symmetry points in the Brillouin zone, the dynamical matrix can be diagonalized analytically.

\subsection{\label{subsec:spin_waves_single_q} Symmetry properties in the decoupled and weak SOC limit}  

Before discussing the spin wave dispersions for the single-$\mathbf{Q}$ structures, it is useful to analyze the symmetry properties of the spin configurations.
We are interested in determining the groups of symmetries that leave the magnetic structures unchanged, \textit{i.e.} the \textit{stabilizer subgroups}. 
The stabilizers allow one to quickly determine the number and degeneracy of modes at a given wavevector $\mathbf{q}$ in the Brillouin zone.

In the decoupled limit, the stabilizers are subgroups of $\mathcal{G}_J$, which was derived in Sec.~\ref{subsec:symmetry_decoupled}. 
Out of all structures listed in Sec.~\ref{subsec:structure_single_q}, $\boldsymbol{\Phi}_0^{(\boldsymbol{\Gamma})}$ (ferromagnetic) and $\boldsymbol{\Phi}_3^{(\boldsymbol{\Gamma})}$ (collinear antiferromagnetic) configurations deserve special attention, since any spin rotation around an axis collinear to the spins leaves the spin configurations unchanged.
Although these configurations are very similar, their symmetry properties turn out to be fundamentally different.
The stabilizer group of the ferromagnetic state contains all lattice symmetries and can therefore be written as a direct product

\begin{equation}
    \mathcal{S}\left(\boldsymbol{\Phi}_0^{(\boldsymbol{\Gamma})}\right) = \mathrm{D}_{6h}^{(L)}\otimes \mathrm{SO}^{(S)}(2).
\end{equation}
On the other hand, the collinear antiferromagnetic state is not invariant under lattice transformations that interchange A and B layers.
\begin{figure*}[t]
    \centering
    \includegraphics[width=0.9\textwidth]{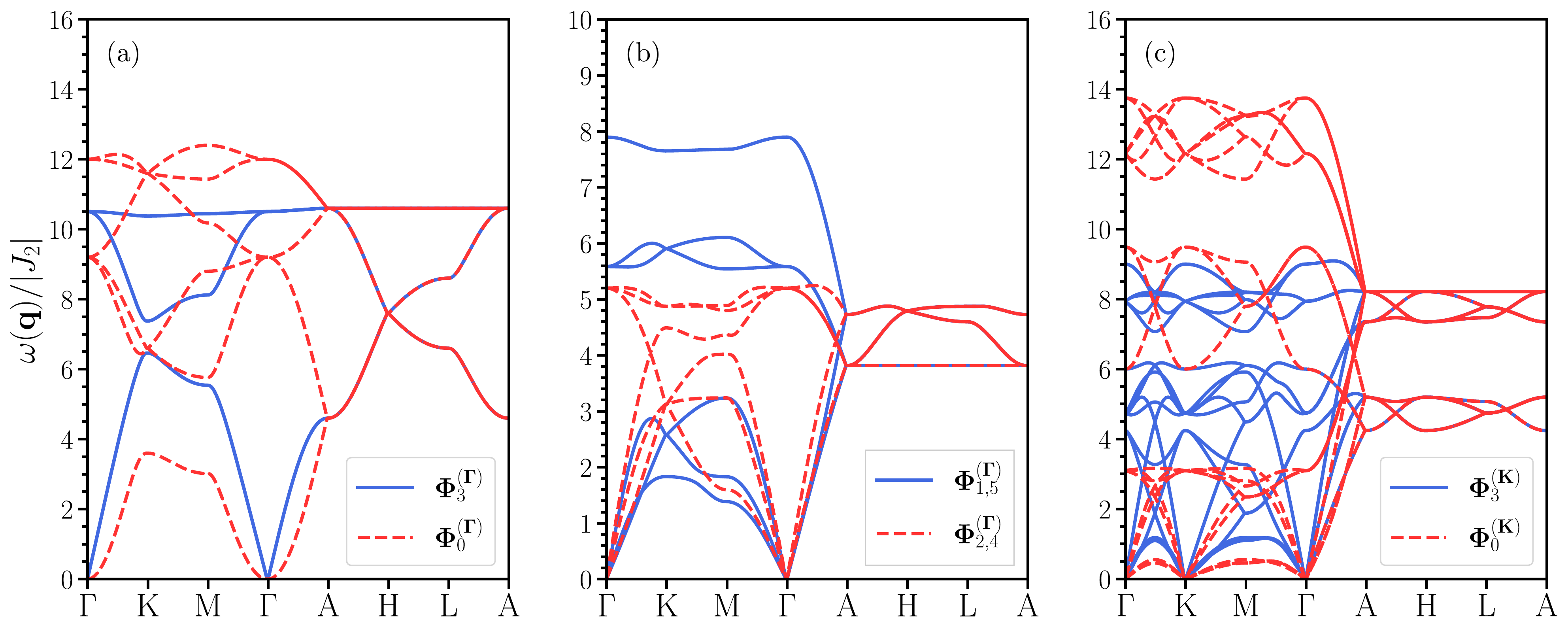}
    \caption{Spin wave dispersions for (a) collinear $\mathbf{Q}=\Gamma$, (b) non-collinear $\mathbf{Q}=\Gamma$, and (c) $\mathbf{Q}=K$ single-$\mathbf{Q}$ structures stabilized in the decoupled limit. In (b), the labels $\boldsymbol{\Phi}^{(\boldsymbol{\Gamma})}_{1,5}$ and $\boldsymbol{\Phi}^{(\boldsymbol{\Gamma})}_{2,4}$ are used to indicate that the phases with $m=1,5$ (2,4) are degenerate and therefore have the same spectra. The parameters used to calculate the dispersions are $J_2<0,J_1/|J_2|=-1,J_4/|J_2|=-0.1$ ($\boldsymbol{\Phi}^{(\boldsymbol{\Gamma})}_{0}$), $J_2<0,J_1/|J_2|=1,J_4/|J_2|=0.1$ ($\boldsymbol{\Phi}^{(\boldsymbol{\Gamma})}_{3}$), $J_2>0,J_1/|J_2|=-1,J_4/|J_2|= 0.1$ ($\boldsymbol{\Phi}^{(\boldsymbol{\Gamma})}_{1,5}$), $J_2>0,J_1/|J_2|=1,J_4/|J_2|=-0.1$ ($\boldsymbol{\Phi}^{(\boldsymbol{\Gamma})}_{2,4}$),  $J_2<0,J_1/|J_2|=-1.5,J_4/|J_2|=1$ ($\boldsymbol{\Phi}^{(\mathbf{K})}_{0}$), and $J_2<0,J_1/|J_2|= 1.5,J_4/|J_2|=-1$ ($\boldsymbol{\Phi}^{(\mathbf{K})}_{3}$).}
    \label{fig:sw_exchange}
\end{figure*}
The remaining symmetry elements form a group $\mathrm{D}_{3h}^{(L)}$, and the stabilizer can therefore be written as a semidirect product:

\begin{equation}
    \mathcal{S}\left(\boldsymbol{\Phi}_3^{(\boldsymbol{\Gamma})}\right) = \mathrm{D}_{3h}^{(L)}\otimes \left[\mathrm{SO}^{(S)}(2)\ltimes \mathrm{C}_2^{(SL)}\right],
\end{equation}
where $\mathrm{C}_2^{(SL)}$ contains a simultaneous $C_2$ rotation of the lattice around $z$-axis and spins around an axis perpendicular to the spins.
This subtle difference in the structure of the stabilizers leads to significant differences in the corresponding spin wave dispersion spectra.
Fig.~\ref{fig:sw_exchange}~(a) demonstrates the differences between the dynamics of these two spin configurations.
The most notable difference is that in the case of $\boldsymbol{\Phi}_3^{(\boldsymbol{\Gamma})}$, all branches are at least doubly degenerate, leading to three distinct modes.
However, as seen in the figure, in the case of a ferromagnetic state, there are generally six non-degenerate modes.
The existence of the two-fold degeneracy in the antiferromagnetic case can be attributed to the fact that at an arbitrary wavevector $\mathbf{q}$ the dynamical matrix in Eq.~(\ref{eq:dynamical_matrix}) is invariant under a simultaneous lattice inversion and reversal of the spin direction.
This situation is equivalent to the spin waves on a linear antiferromagnetic chain where two linearly-polarized magnon modes have the same dispersion.  
As a result, the collinear antiferromagnetic state has two linear Goldstone modes at the $\Gamma$ point, corresponding to the in- and out-of-phase fluctuations of spins on A and B sublattices.
In contrast, the ferromagnetic state only has one (quadratic) Goldstone mode.

In the case of the non-collinear single-$\mathbf{Q}$ configurations, the symmetry analysis is simplified by the fact that the stabilizer subgroups are now finite groups, since there are no spin rotations that leave these structures invariant.
One can see from Fig.~\ref{fig:q0_phases} that for each non-collinear $\boldsymbol{\Phi}_m^{(\boldsymbol{\Gamma})}$ configuration there is exactly one spin rotation that can be combined with a given lattice symmetry transformation to leave the spin configuration unchanged.
Therefore, the stabilizer subgroups for $\mathbf{Q}=\Gamma$ non-collinear phases are all isomorphic to $\mathrm{D}_{6h}$.
The situation is similar in the case of $\boldsymbol{\Phi}_m^{(\mathbf{K})}$ structures.
However, because certain lattice symmetries are broken as a result of the cycloidal spin order, we are left with the symmetry operations that belong to the group of the ordering wavevector $\mathbf{Q}=K$.
Thus, the stabilizers of $\boldsymbol{\Phi}_m^{(\mathbf{K})}$ are all isomorphic to group $\mathrm{D}_{3h}$.
Figures~\ref{fig:sw_exchange}~(b-c) show the dispersions for the non-collinear single-$\mathbf{Q}$ spin configurations in phase diagrams in Fig.~\ref{fig:pd_j1_j4_h}.
Note that the non-collinear nature of these phases implies that the Goldstone modes are three-fold degenerate, since there are now three non-equivalent fluctuation axes.

In the weakly-coupled limit, the stabilizers must be the subgroups of $\mathcal{G}_D$.
This does not change the symmetry of the non-collinear single-$\mathbf{Q}$ states, although DM interactions lift the three-fold degeneracy of the Goldstone modes, leaving only one zero-energy mode.
Since the out-of-plane spin rotations are no longer valid symmetry operations, the stabilizers of the two collinear phases discussed above formally become isomorphic to $\mathrm{D}_{6h}$, similar to the rest of the $\boldsymbol{\Phi}_m^{(\boldsymbol{\Gamma})}$ configurations.
We note that when one considers the symmetry of the local spin components used for spin-wave calculations, the stabilizer groups become the same (\textit{i.e.} not just isomorphic).
Therefore, the spin wave eigenvectors in the local frame are exactly the same for all six $\boldsymbol{\Phi}_m^{(\boldsymbol{\Gamma})}$ ($\boldsymbol{\Phi}_m^{(\mathbf{K})}$) phases.
This important property holds true for all magnetic structures in the weak SOC limit and is one of the consequence of the self-duality discussed in Sec.~\ref{sec:self_duality}.

\subsection{\label{subsec:spin_waves_obdo_norec} Order by disorder and non-reciprocity}  

\begin{figure}[t]
    \centering
    \includegraphics[width=0.5\textwidth]{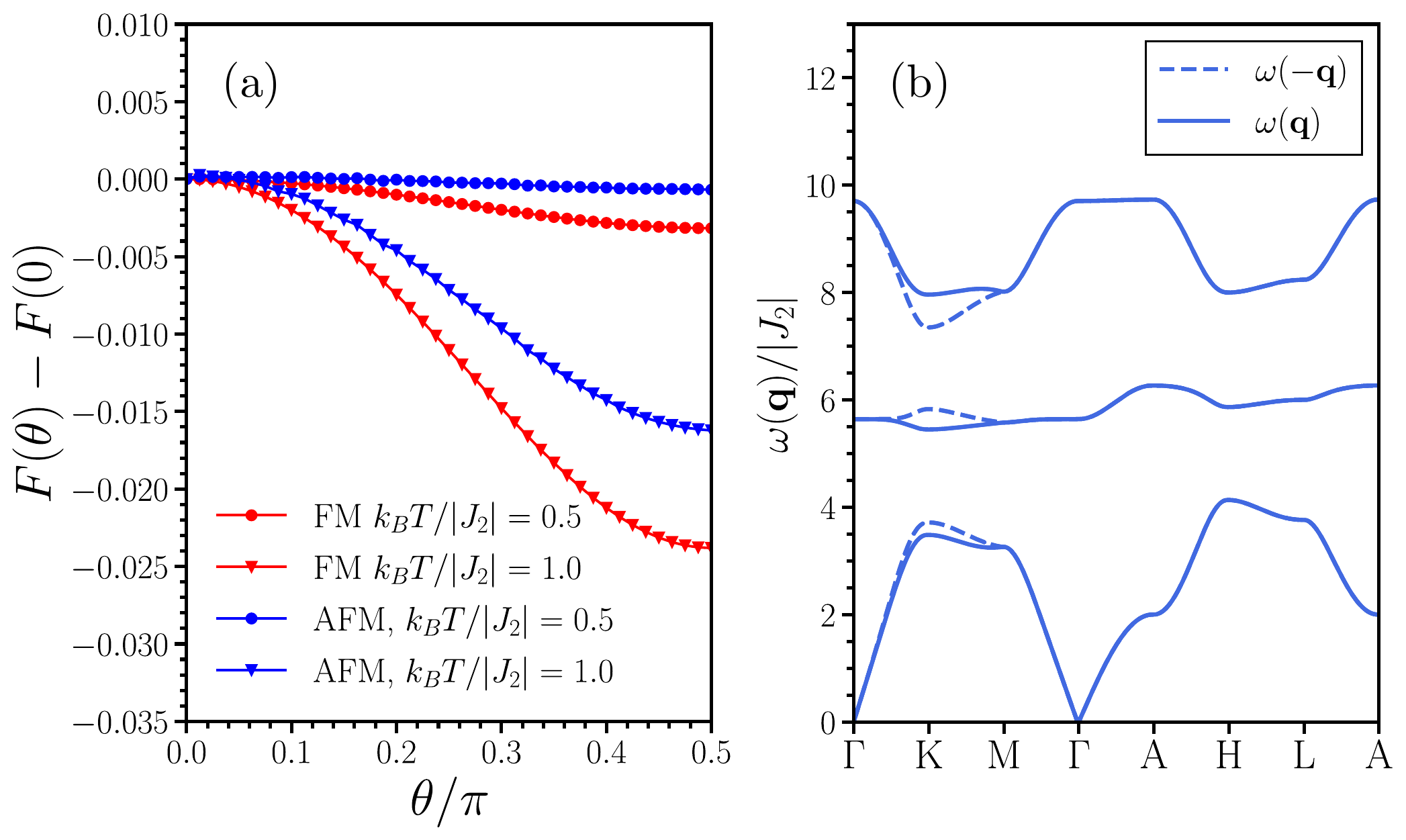}
    \caption{(a) Magnon free energy as function of the out-of-plane angle $\theta$ calculated numerically for the two collinear phases ($\boldsymbol{\Phi}_{0,3}^{(\boldsymbol{\Gamma})}$) with $J_2<0$, $J_1/|J_2|=0.5$, $D_1=D_2=0.5$. The minima of the free energy occur at $\theta=\frac{\pi}{2}$, indicating order by disorder. (b) Non-reciprocity in the spin wave spectrum of the antiferromagnetic collinear state in the weak SOC limit.}
    \label{fig:obdo_nonrec}
\end{figure}

As discussed in Sec.~\ref{subsec:structure_single_q}, the classical energy of the collinear phases does not depend on the DM constants, meaning that these spin configurations can be rotated out-of-plane without any energy cost.
Nevertheless, the energy of the spin waves does, in fact, depend on the DM interactions.
As a result, we expect that the fluctuations would break the effective rotational symmetry by selecting a particular orientation of the collinear structures via the order-by-disorder mechanism~\cite{Henley_obdo_1989_prl,Chalker_Holdsworth_Shender_kagome_1992_prl}.
To prove demonstrate predictions, we calculate the magnon free energy as function of the out-of-plane angle $\theta$

\begin{equation}
    F(\theta) = \frac{1}{\beta}\sum_\mathbf{q} \ln\left(1-e^{-\beta\omega(\mathbf{q},\theta)}\right).
\end{equation}
The calculations of the frequencies $\omega(\mathbf{q},\theta)$ were performed numerically using a $50\times 50\times 50$ discretized reciprocal lattice grid. 
The result is shown in Fig.~\ref{fig:obdo_nonrec}~(a).
We can see that in both cases, thermal fluctuations rotate the spins perpendicular to the kagome layers, thus preserving the
continuous rotational symmetry in these phases. 

We note that in the case of the collinear antiferromagnet, the stabilization of the out-of-plane order through thermal fluctuations introduces an interesting new property in the spin wave spectrum, namely non-reciprocity~\cite{Rikken_Strohm_Wyder_nonrec_2002_prl,Zakeri_nonrec_2010_prl,Garcia-Sanchez_nonrec_2014_prb,Cheon_Lee_Cheong_norec_2018_prb,Santos_nonrec_2020_prb,Borys_Monchesky_nonrec_2021_prb}.
Non-reciprocity implies that there is no unitary transformation that could make the dynamical matrix~(\ref{eq:dynamical_matrix}) Hermitian.
As a result, the spectrum is no longer symmetric at $\omega(\pm\mathbf{q})$, as seen in Fig.~\ref{fig:obdo_nonrec}~(b).

Finally, we note that the $\boldsymbol{\Phi}_{0,3}^{(\mathbf{K})}$ configurations also possess the same accidental rotational degeneracy.
This is evidenced from the fact that the classical energy of these configurations is independent of the DM constants.
However, because the stabilizer groups of these phases are significantly smaller than those of the collinear phases, the out-of-plane rotations in this case generally lower the symmetry of the spin configurations.
Instead, the breaking of the continuous degeneracy occurs as a result of small spin deviations that establish finite DM coupling.

\subsection{\label{subsec:spin_waves_multi_q}Excitations in the multi-$\mathbf{Q}$ phases}

Next, we calculate the excitation spectra for the $\boldsymbol{\Psi}_m^{(\mathbf{M})}$ phases.
Unlike in the case of the $\mathbf{Q}=K$ single-$\mathbf{Q}$ states, it is not possible to obtain all of the spin wave modes by simply shifting the branches by $\pm K$, and one has to construct a dynamical matrix where each of the four elements $u_{ij}^{(ab)}$ is a $24\times 24$ matrix (as the magnetic unit cell consists of 4 crystallographic unit cells).
As a result, the excitation spectra for these states consist of 24 modes.
\begin{figure}[t]
    \centering
    \includegraphics[width=0.5\textwidth]{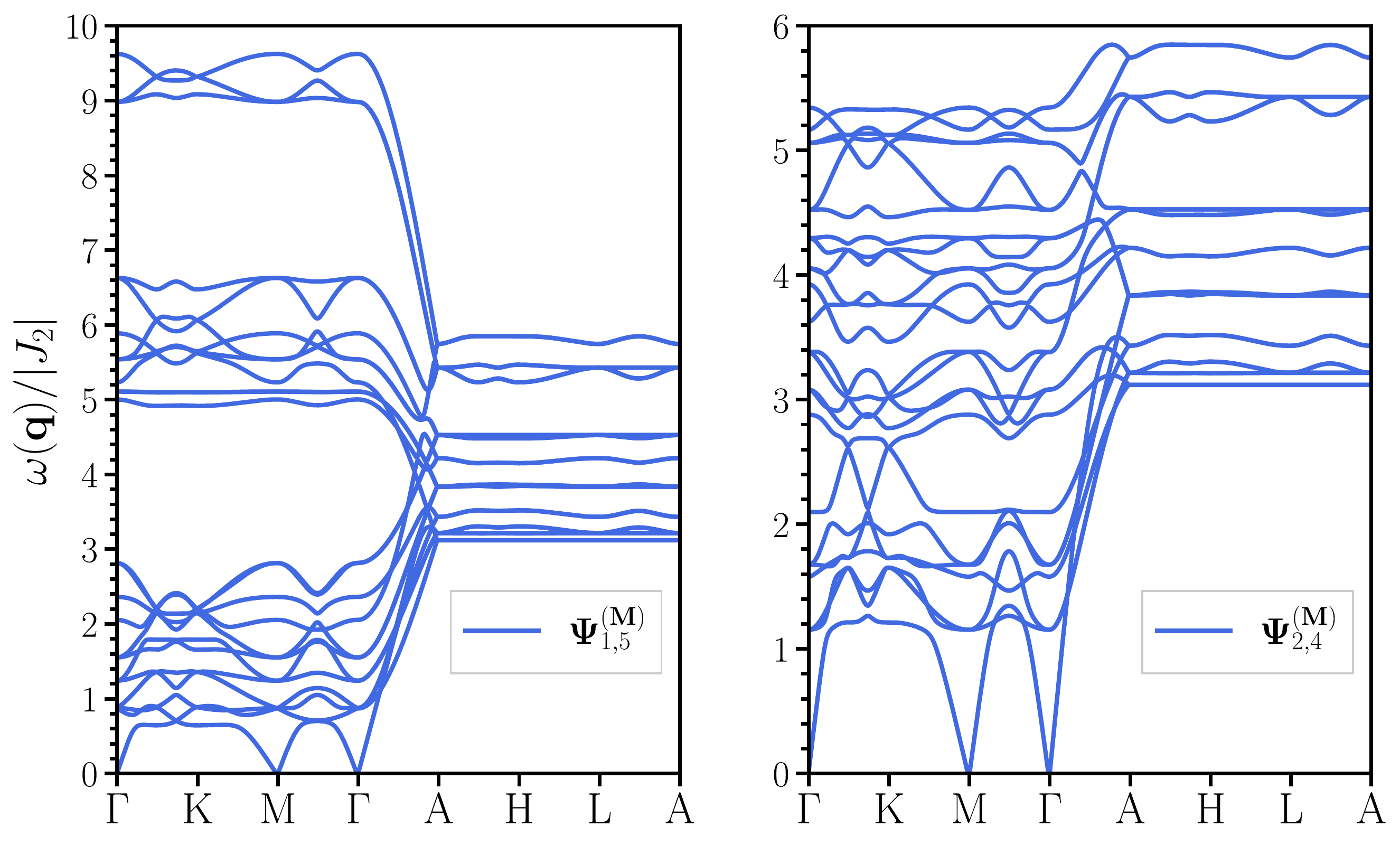}
    \caption{Spin wave dispersion for the four out of six  $\boldsymbol{\Psi}_m^{(\mathbf{M})}$ configurations. The parameters used for the calculation are $J_2>0$, $J_1/|J_2|=1$, $D_1/|J_2|=\mp 0.5$, $D_2/|J_2|=\mp 0.5$ for $\boldsymbol{\Psi}_{2,4}^{(\mathbf{M})}$, and $J_2>0$, $J_1/|J_2|=-1$, $D_1/|J_2|=\pm 0.5$, $D_2/|J_2|=\mp 0.5$ for $\boldsymbol{\Psi}_{1,5}^{(\mathbf{M})}$.}
    \label{fig:sw_4q}
\end{figure}
Fig.~\ref{fig:sw_4q} shows the dispersions for the $\boldsymbol{\Psi}_m^{(\mathbf{M})}$ phases stabilized for $J_2>0$, $|J_1|=|J_2|$ in the weak SOC limit.
As in the case of the single-$\mathbf{Q}$ states, the DM interactions lead to a single Goldstone mode, which in this case occurs at $\Gamma$ and M points, since the period of the magnetic texture is double that of the lattice.
Certain branches in the spectra also exhibit other interesting features such as flat bands. However further studies are needed for a full description of band topology in these phases.

\begin{figure*}
    \centering
    \includegraphics[width=\textwidth]{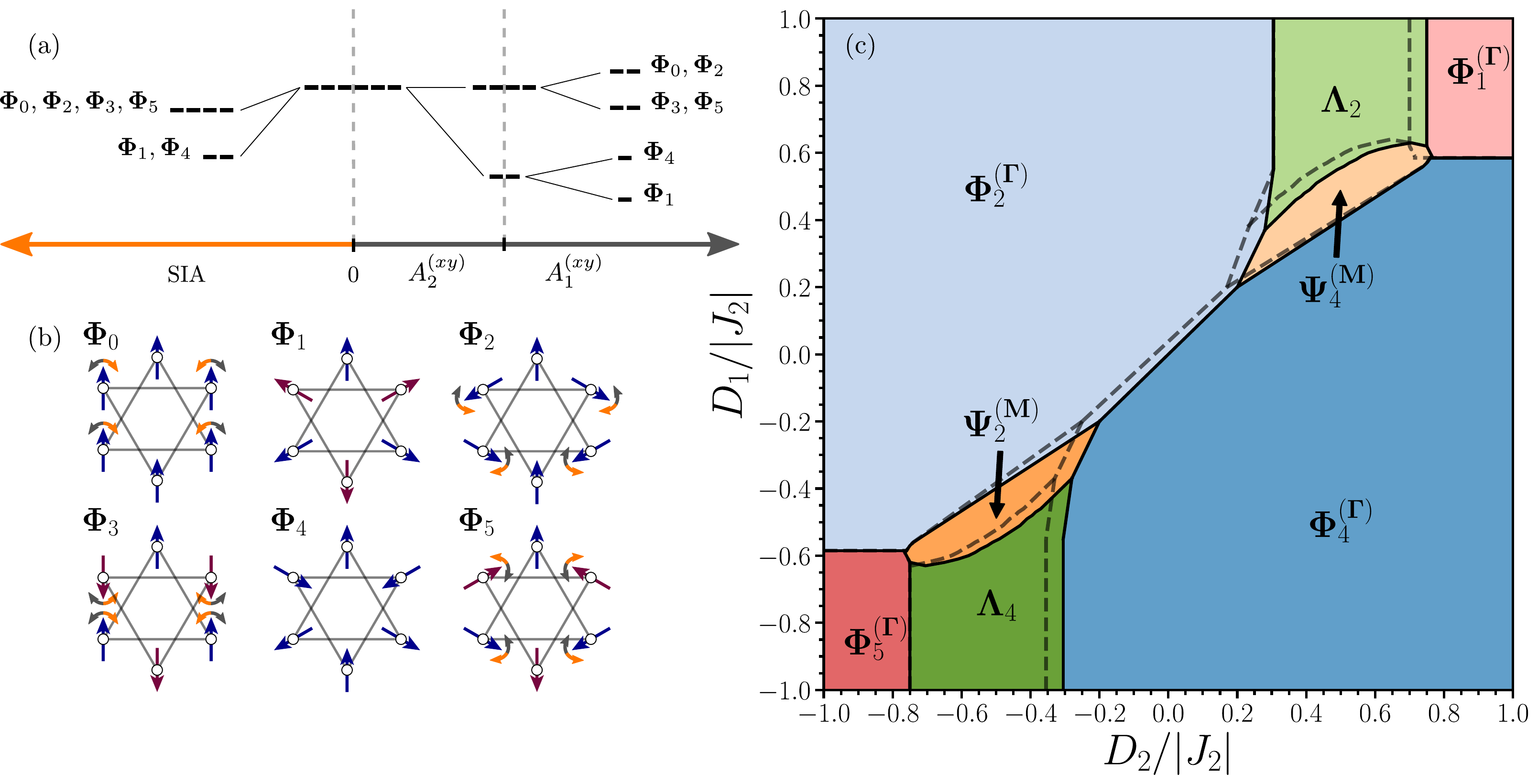}
    \caption{(a) A diagram illustrating the splitting  of the classical energies of the six planar irreps in the weak SOC limit. Here, the values of the exchange and DM constants are chosen using the self-duality relations to give the same energies in the zero anisotropy limit. Note that the energy scales are not exact. (b) Distortion of the spin structures as a result of small SIA (orange arrows) and bond-dependent anisotropy (grey arrows). The $\boldsymbol{\Phi}_{1,4}^{(\boldsymbol{\Gamma})}$ remain unchanged since the spins are collinear with the anisotropy axes. (c) Phase diagram in Fig.~\ref{fig:pds} (left) with shifted phase boundaries (grey dashed lines) introduced by a small SIA ($K^-/|J_2|= 0.05$).}
    \label{fig:anis_effects}
\end{figure*}

\section{\label{sec:anisotropy}Effects of anisotropic interactions}

\subsection{\label{subsec:anisotropy_structure} Phase stability}

Having described the important properties of the magnetic phases in the decoupled and weak SOC limits, we now consider the case of the intermediate SOC.
We assume that in this limit, the SIA and bond-dependent anisotropy are small but non-negligible.

As discussed in Sec.~\ref{sec:symmetry}, the anisotropic interactions break the axial rotational symmetry, reducing the symmetry group of the Hamiltonian to the paramagnetic group.
As a result, the $\boldsymbol{\Phi}_0^{(\boldsymbol{\Gamma})}$ ($\boldsymbol{\Phi}_3^{(\boldsymbol{\Gamma})}$) and $\boldsymbol{\Phi}_2^{(\boldsymbol{\Gamma})}$ ($\boldsymbol{\Phi}_5^{(\boldsymbol{\Gamma})}$) states now belong to the same two-dimensional irreps, while the remaining $\boldsymbol{\Phi}_1^{(\boldsymbol{\Gamma})}$ and $\boldsymbol{\Phi}_4^{(\boldsymbol{\Gamma})}$ split into one-dimensional irreps (see Fig.~\ref{fig:SOC_symmetry_diagram}). 
In general, the anisotropic interactions stabilize the spin configurations that belong to the one-dimensional planar irreps, since in these structures the spins are aligned collinear to the anisotropy axes.
At the same time, the configurations belonging to the same two-dimensional planar irreps remain almost degenerate for small values of the anisotropy constants.
This is illustrated in Fig.~\ref{fig:anis_effects}~(a).
We select the exchange and DM parameters using self-duality relations in Sec.~\ref{subsec:duality_JD} such that in the absence of anisotropy the energies of the six planar $\boldsymbol{\Phi}_m^{(\boldsymbol{\Gamma})}$ configurations are the same.
The SIA and the in-plane bond-dependent interactions split the energies by stabilizing $\boldsymbol{\Phi}_1^{(\boldsymbol{\Gamma})}$ and $\boldsymbol{\Phi}_4^{(\boldsymbol{\Gamma})}$ (the sign of $K^-$ and $A_2^{(xy)}$ determines the orientations of spins).
The out-of-plane bond anisotropy further splits the energies based on the parity of the configurations (in the figure, we choose $A_1^{(xy)}>0$, which stabilizes the structures that are antisymmetric under space inversion).

Apart from the energy of the spin configurations, anisotropic interactions have a significant impact on the spin structures themselves.
Since some of the planar phases now belong to the same two-dimensional irreps, the anisotropic interactions couple these states, introducing small deviations from the irreps in the weak SOC limit, as shown in Fig.~\ref{fig:anis_effects}~(b).
This has been shown to play an important effect on the physical properties of $\mathrm{Mn}_3X$ compounds, where the anisotropic interactions couple the $\boldsymbol{\Phi}_2^{(\boldsymbol{\Gamma})}$ 120-degree state and the ferromagnetic state, inducing a small magnetic moment at zero field~\cite{Soh_gs_2020_prb,Chen_gs_2020_prb,Zelenskiy_Monchesky_Plumer_Southern_2021_prb}.
Our previous work in Ref.~I has established that the relative strengths of the SIA and the in-plane bond-anisotropy fix the overall orientation of the spins as well as the direction of the spin deviations, consequently determining both the magnitude and direction of the induced magnetic moment in $\mathrm{Mn}_3X$ magnets.
\begin{figure*}[t]
    \centering
    \includegraphics[width=\textwidth]{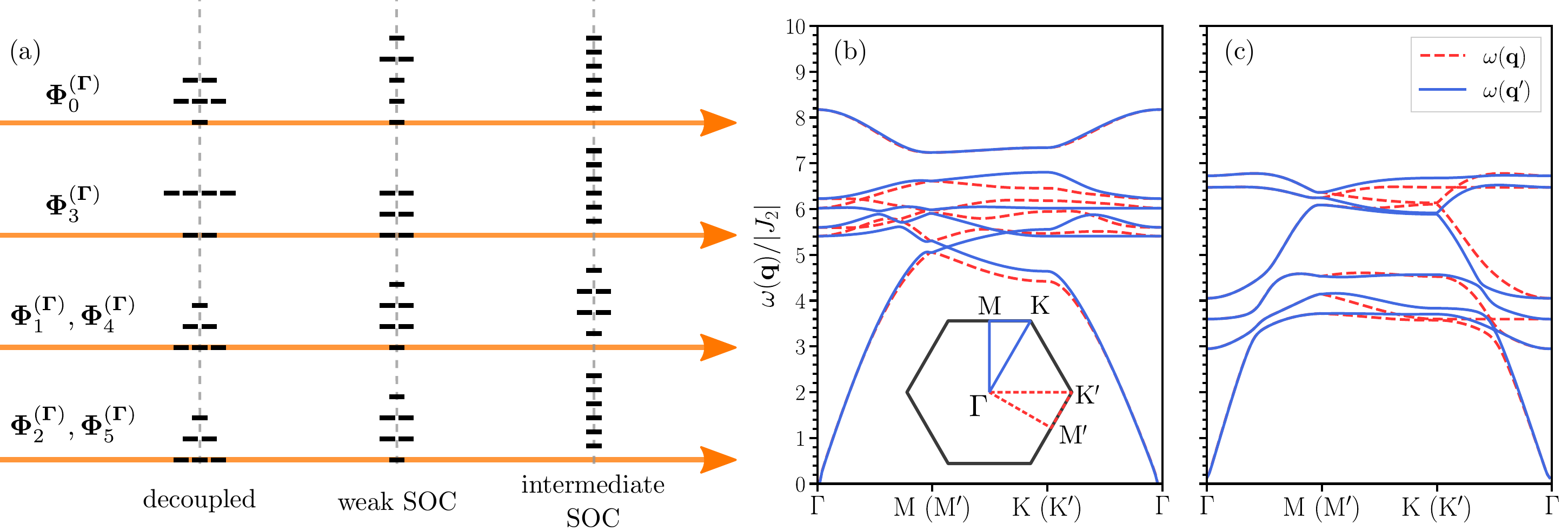}
    \caption{(a) Diagram demonstrating the $\Gamma$-point splitting of the spin wave modes in $\boldsymbol{\Phi}_m^{(\boldsymbol{\Gamma})}$ phases in the three SOC limits. Examples of broken symmetry in the excitations of (b) $\boldsymbol{\Phi}_2^{(\boldsymbol{\Gamma})}$ and (c) $\boldsymbol{\Phi}_5^{(\boldsymbol{\Gamma})}$. The solid blue and dashed red lines correspond to dispersions along the two paths illustrated in the inset of figure (b). In both cases, we set $K^-/|J_2|=0.1$, $J_2>0$, $J_1/|J_2|=1$, $D_2/|J_2|=-0.8$, and set $D_1/|J_2|=0.8$ for $\boldsymbol{\Phi}_2^{(\boldsymbol{\Gamma})}$, and $D_1/|J_2|=-0.8$ for $\boldsymbol{\Phi}_5^{(\boldsymbol{\Gamma})}$.}
    \label{fig:SOC_spin_waves}
\end{figure*}
We note that the distortions of the spin structures introduced by the anisotropic couplings may also provide a way for the experimental quantification of these interactions.
Techniques, such as elastic neutron scattering, would allow one to determine the canting angles, which could in turn be related to the values of the anisotropic parameters.
Moreover, in the case of the coupling between the ferromagnetic $\boldsymbol{\Phi}_0^{(\boldsymbol{\Gamma})}$ state and $\boldsymbol{\Phi}_2^{(\boldsymbol{\Gamma})}$, the induced magnetic moment would serve as an excellent probe of the anisotropy in the system. 

Although the discussion in this section so far has been in the context of the $\mathbf{Q}=\Gamma$ phases, the general trends in the stabilization energy and spin configurations mostly apply to the remaining types of phases, since every spin structure discussed in this paper can be approximately constructed from the rotated $\boldsymbol{\Phi}_m^{(\boldsymbol{\Gamma})}$ states.
As the deviations from the $\boldsymbol{\Phi}_m^{(\boldsymbol{\Gamma})}$ local order parameters are small, we are still able to make qualitative predictions about the stability of different phases under small anisotropy.
Therefore, we expect the anisotropic interactions to stabilize those magnetic phases characterized by $m=1,4$, thus extending their stability regions as compared to the weak SOC limit.
This is clearly demonstrated in Fig.~\ref{fig:anis_effects}~(c), where the phase boundaries are calculated using $J_2>0$, $J_1/|J_2|=1$, $K^-/|J_2|=0.05$.
The symmetry of the phase diagram is broken, since $\mu_0^{(-1)}$ is no longer a valid self-duality transformation.
Notably, the in-plane bond-dependent anisotropy with $A_2^{(xy)}=0.05$ produces a nearly identical shift of the phase boundaries.
The only exception to the established stability trend appears to be the $\boldsymbol{\Lambda}_4$ phase which shrinks under the applied anisotropy.
However, this is not surprising given the result of Sec.~\ref{subsec:structure_ising} where the spin deviations were shown to be the primary stabilizing factor of the Ising patterns.  

\subsection{\label{subsec:anisotropy_spin_waves}Spin wave spectra}
\begin{table}[!t]
    \centering
    \begin{tabular}{c|c|c}
        m & Stabilizer & irrep decomposition at $\Gamma$\\
        \hline
        0 & C$_{2h}^{(SL)}$            & $A_g\oplus A_u\oplus 2B_g\oplus 2B_u$\\
        1 & D$_{6 }^{(SL)}$ & $A_2\oplus B_1\oplus E_1\oplus E_2$\\
        2 & C$_{2h}^{(SL)}$            & $A_g\oplus A_u\oplus 2B_g\oplus 2B_u$\\
        3 & D$_{2 }^{(SL)}$ & $A_1\oplus 2B_1\oplus B_2\oplus 2B_3$\\
        4 & D$_{3d}^{(SL)}$            & $2A_2''\oplus 2E''$\\
        5 & D$_{2 }^{(SL)}$ & $A_1\oplus 2B_1\oplus B_2\oplus 2B_3$\\
    \end{tabular}
    \caption{Stabilizer groups and irrep decomposition at the $\Gamma$ point for the six $\boldsymbol{\Phi}_m^{(\boldsymbol{\Gamma})}$ states in the intermediate anisotropy limit.}
    \label{tab:anisotropy_splitting}
\end{table}
The effects of anisotropic interactions are strongly manifested in the excitations of the magnetic structures.
Broken rotational symmetry implies that the spin wave spectra are now completely gapped.
Since the symmetry group of the Hamiltonian is reduced down to the paramagnetic group, the stabilizer subgroups of the spin configurations in the intermediate SOC limit are typically small.
In the case of the six $\mathbf{Q}=\Gamma$ phases, the stabilizer groups and the corresponding irrep decompositions at the $\Gamma$ point are presented in table~\ref{tab:anisotropy_splitting}.
We see that the stabilizers of states with $m=0,2$ ($\mathrm{C}_{2h}$) and $m=3,5$ ($\mathrm{D}_2$) contain only one-dimensional irreps, meaning that apart from the accidental mode crossings, the spin wave spectra will generally only contain non-degenerate modes.
However, the remaining two configurations retain most of their symmetry from the weak SOC limit and therefore have doubly-degenerate modes. 

We now briefly summarize the effects of the SOC strength on the spin wave spectra.
Fig.~\ref{fig:SOC_spin_waves}~(a) demonstrates the evolution of the $\Gamma$ point splitting of excitations in the six $\boldsymbol{\Phi}_m^{(\boldsymbol{\Gamma})}$ phases.
The transitions between the different SOC limits are accompanied by a qualitative change in the spectra, associated with either a change in the splitting or an opening of a gap.
In addition, since the spin structures break most of the symmetries of the Hamiltonian, the spectra in general will not be symmetric in the Brillouin zone.
This is demonstrated in Fig.~\ref{fig:SOC_spin_waves}~(b) and (c), where two different paths in the Brillouin zone, related by a reflection yield different values of $\omega(\mathbf{q})$.

These qualitative changes in the spin wave spectra provide a good probe for the SOC strength in the AB-stacked kagome magnets.
By comparing these results to, for example, the inelastic neutron scattering or Raman scattering data, one would be able to determine the appropriate SOC limit, and thus identify the relevant spin interactions. 

\section{\label{sec:conclusions} Summary and Conclusions}

In this work, we provide an extensive overview of the properties of the general magnetic Hamiltonian~(\ref{eq:magnetic_hamiltonian}) on hexagonal AB-stacked lattices.
By studying the symmetry of the model, we have determined the connection between the strength of the SOC and the allowed spin symmetries in the system.
We have further identified three cases corresponding to decoupled, weak, and intermediate SOC limits that yield different Hamiltonian symmetry groups.
In addition to the symmetries, we found a large number of self-duality transformations that define the structure of the parameter space of the model.
Since these transformations directly depend on the symmetry of the physical system, we identify three sets of duality transformations corresponding to the three SOC limits.

The fundamental properties of the Hamiltonian allowed us to devise a strategy for efficient exploration of the parameter space.
We studied the ordered phases in the decoupled and weak SOC limits, by constructing parameter-space phase diagrams, using a combination of analytical LT and numerical MC methods. 
We analyzed the structures of the resulting spin configurations, including single- and multi-$\mathbf{Q}$ phases, and gave exact or nearly exact parameterizations.
Among the most interesting structures identified in this work are the Ising-like patterns found in large pockets of the parameter space in the weak and intermediate SOC limits.
We determine that these states are stabilized by small deviations from the idealized order parameters, and derive the second order solution for the optimal spin canting, as well as the stabilization energy.

Next, we calculated the spin-wave spectra of some of the single- and multi-$\mathbf{Q}$ phases and analyzed the symmetry of the spin fluctuations.
We found that in the weak SOC limit the fluctuations drive the collinear configurations out of the kagome plane, signifying order-by-disorder.
As a consequence of this rotation, the DM interactions make the excitations in the collinear antiferromagnet phase non-reciprocal.

Finally, we study the effects of the intermediate SOC, manifested by the additional SIA and bond-dependent interactions. 
We find that small amounts of anisotropy can produce traceable changes in the structure and excitations of the spin configurations, which opens the doors for potential experimental quantification of these interactions.

As discussed throughout the paper, the results of this work are of direct relevance to the $\mathrm{Mn}_3X$ family of compounds.
For example, the interplay of the anisotropic interactions has already been shown in our previous work~Ref.~I to produce detectable changes in the static and dynamic properties of these compounds.
Another promising magnetic compound with the AB-layered structure is the $\mathrm{Gd}_3\mathrm{Ru}_4\mathrm{Al}_{12}$~\cite{Chandragiri_Gd_2016_IOP,Matsumura_Gd_2019_jpsj,Hirschberger_Gd_2019_nature_comm}. 
Although the magnetic parameters for these materials may not fall in any of the non-trivial phases (such as multi-$\mathbf{Q}$ structures), one may still be able to realize these states in systems where the SOC can be tuned experimentally.
For example, in thin-layer systems, the SOC close to the surface can be enhanced by an addition of a heavy-metal capping layer~\cite{Gradmann:1986jmmm,Heinrich:1998}.
Importantly, this study has shown that the unconventional phases often lie within the typical physical range of magnetic parameters, including the DM and anisotropic couplings.
In addition, the existence of self-duality maps significantly increases the chances that a compound with AB-layered kagome structure would exhibit non-trivial magnetic phenomena.
The last statement can be justified by recalling that in Sec.~\ref{subsec:structure_ising} the $\boldsymbol{\Lambda}$ phase was related via duality to a model with large ferromagnetic out-of-plane exchange ($J_1\ll 0$), strong in-plane DM interactions ($D_2\neq 0$) and negligible in-plane exchange ($J_2\approx 0$).
Under normal circumstances, this situation would be extremely difficult to realize, rendering the model as unphysical.
However, the duality provides us with a dual images that lie within a reasonable range of parameters.

In conclusion, we would like to point out that some of the assumptions made in the beginning of this article could have an important impact on some of the magnetic properties of $\mathrm{Mn}_3X$ systems.
Since the interactions are predominantly governed by the itinerant electrons, long-range interactions may actually play an important role in stabilizing certain experimentally observed phases.
For example, further neighbor interactions along the $\mathbf{\hat{z}}$-direction would be required in order to stabilize out-of-plane spatial modulations through competition with the nn couplings~\cite{Zelenskiy_Rb_2021_prb}.
This also means that the breathing anisotropy may also affect certain properties of the magnetic phases.
Furthermore, some recent works have been dedicated to the importance of the coupling between the elastic and magnetic degrees of freedom~\cite{Theuss_strain_2022_prb,Reis_strain_2020_prm,Sukhanov_strain_2018_prb,Sukhanov_strain_2019_prb}. 
These are known to introduce effective biquadratic spin interactions, which would further complicate the theoretical analysis.
As it stands however, this work provides an invitation for further investigations of the rich magnetic phenomena offered by the AB-stacked kagome compounds.

\section{\label{sec:acknowledgements} Acknowledgements}
This work was supported by the Natural Sciences and Engineering Research Council of Canada (NSERC).
The authors would like to thank J. S. R. McCoombs, C Rudderham, D. Kalliecharan, and B. D. MacNeil for helpful discussions.
We acknowledge D. Kalliecharan's suggestions for the visual presentation of this paper.
Finally, we thank S. H. Curnoe and J. P. F. LeBlanc for their useful comments in regards to the symmetry and self-duality sections.

\end{document}